\documentclass[aps,preprint,nofootinbib,superscriptaddress]{revtex4}%
\usepackage{amssymb}
\usepackage{amsmath}
\usepackage{epsfig}
\usepackage{amsfonts}
\usepackage{graphicx}
\usepackage{hyperref}%
\providecommand{\U}[1]{\protect\rule{.1in}{.1in}}
\begin{document}
	\title{Extended Complex Yang-Mills Instanton Sheaves}
	\author{Sheng-Hong Lai}
	\email{xgcj944137@gmail.com}
	\affiliation{Department of Electrophysics, National Chiao-Tung University, Hsinchu, Taiwan, R.O.C.}
	\author{Jen-Chi Lee}
	\email{jcclee@cc.nctu.edu.tw}
	\affiliation{Department of Electrophysics, National Chiao-Tung University, Hsinchu, Taiwan, R.O.C.}
	\author{I-Hsun Tsai}
	\email{ihtsai@math.ntu.edu.tw}
	\affiliation{Department of Mathematics, National Taiwan University, Taipei, Taiwan, R.O.C. }
	\date{\today}
	
	\begin{abstract}
		In the search of YM instanton sheaves with topological charge two, the rank of
		$\beta$ matrix in the monad construction can be dropped from the bundle case
		with rank $\beta$= $2$ to either rank $\beta$= $1$ [4] or $0$ on some points
		of $CP^{3}$ of the sheaf cases$.$ In this paper, we first show that the sheaf
		case with rank $\beta$= $0$ does not exist for the previous construction of
		$SU(2)$ complex YM instantons [3]. We then show that in the new "extended
		complex YM instantons" discovered in this paper, rank $\beta$ can be either
		$2$ on the whole $CP^{3}$ (bundle) with some given ADHM data or $1,$ $0$ on
		some points of $CP^{3}$ with other ADHM data (sheaves). These extended $SU(2)$
		complex YM instantons have no real instanton counterparts.
		
	\end{abstract}
	\maketitle
	\tableofcontents

	%
	
	\setcounter{equation}{0}
	\renewcommand{\theequation}{\arabic{section}.\arabic{equation}}%

	\section{Introduction}
	
	Recent developments of complex Yang-Mills (YM) instantons
	\cite{math2,math3,Ann,Ann2,Ann3,Ann4} have revealed many new mathematical
	structures of YM instantons which were not realized in the previous real YM
	instantons \cite{U(1),the,BPST,CFTW,JR,Ward,AW,ADHM,CFYG,CSW,JR2}. In
	particular, for the case of $SU(2)$ complex YM instantons \ (or non-compact
	$SL(2,C)$ \cite{WY} real YM instantons \cite{Lee,math4}), it was shown that
	for some ADHM data at some points on $CP^{3}$ the vector bundle description of
	$2$-instanton breaks down, and one is led to use a description in terms of
	torsion free sheaves for these "instanton sheaves" on $CP^{3}$ \cite{Ann2}.
	The existence of these $2$-instanton sheaves was soon extended to the complex
	$k$-instanton sheaves with higher topological charges $k=3,4$ \cite{Ann3}.
	
	On the other hand, one unexpected result obtained in \cite{Ann4} was the great
	simplification or "solvability" of calculation of the field strength $F$
	associated with the sheaf ADHM data, and the explicit form of a class of
	$SU(2)$ complex YM $2$-instanton field strength without removable
	singularities can be exactly calculated ! This "solvability" was not available
	for the previous real YM $k$-instanton calculation with $k\geq2$, and was
	presumably related to the imposing of the sheaf condition \cite{math2,math3}
	on YM instanton sheaves.
	
	Moreover, one sheaf line which is the real line supporting sheaf points on
	$CP^{3}$ of $SU(2)$ YM $2$-instanton sheaf was identified and found to be a
	special jumping line over $S^{4}$ spacetime \cite{Ann4}. In addition, the
	order of the singularity structure of the connection $A$ and the field
	strength $F$ at the corresponding singular point on $S^{4}$ associated with
	this sheaf line in $CP^{3}$ was found to be higher than those of other
	singular points associated with the normal jumping lines \cite{Ann4}. These
	results suggest that the existence of YM instanton sheaves on $CP^{3}$ is
	closely related to the new singular structure of the corresponding $A$ and $F$
	on $S^{4}$ spacetime.
	
	In the search of YM instanton sheaves with topological charge two \cite{Ann2},
	the rank of $\beta$ matrix in the monad construction can be dropped from $2$
	of the bundle case to either $1$ presented in \cite{Ann2} or $0$ on some
	points of $CP^{3}$ of the sheaf cases$.$ In this paper, we will first show
	that the sheaf case of rank $\beta$= $0$ does not exist for the previous
	construction of $SU(2)$ complex YM instantons ($\det\neq0$). We then show that
	in the new "extended complex YM instantons" ($\det=0$) discovered in this
	paper, the rank of $\beta$ can be either $2$ on the whole $CP^{3}$ with some
	given ADHM data (bundle) or $1,$ $0$ on some points of $CP^{3}$ with other
	ADHM data (sheaves). We will see that these new extended $SU(2)$ complex YM
	instantons have no real instanton counterparts. See Eq.(\ref{poss}) and
	Eq.(\ref{poss2}) in section III for details.
	
	One interesting result we obtained in the search of extended complex YM
	instantons was the discovery of the existence of instanton sheaf structure
	with diagonal $y$ ADHM data. For the $\det\neq0$ complex YM instantons
	constructed previously \cite{Ann}, it was shown \cite{Ann2} that there was no
	instanton sheaf solutions with diagonal $y$ ADHM data.\bigskip\ Recent results
	of new real YM instantons on some special backgrounds can be found at
	\cite{kahler,popov}.
	
	The new complex YM $2$-instanton solutions discovered in this paper do not
	increase the number of moduli parameters $16k-6$ ($k=2$) of the previous
	construction \cite{Ann}. They are "limiting solutions" living on the boundary
	of the moduli space. We expect the existence of these new extended complex YM
	instanton solutions for general topological charge $k$. Recently, some of
	irreducible components of the boundary of the moduli space of instanton
	sheaves of charge $k$ on $CP^{3}$ have been constructed in \cite{math5,math6}
	by mathematicians. In this paper, we will give some explicit constructions for
	$k=2$ case which will be more accessible to physicists. The relation of the
	construction in this paper to the results in \cite{math5,math6} remains to be studied.
	
	This paper is organized as following. In section II, we review the first
	complex YM $2$-instanton sheaves we discovered in \cite{Ann2}. These complex
	YM $2$-instanton sheaves are of rank$\beta=1$ and $\det\neq0$. In section III,
	we show that there is no complex YM $2$-instanton sheaves with rank$\beta=0$
	and $\det\neq0$. The results of section III led us to consider the complex YM
	$2$-instanton sheaves with $\det=0$ which were calculated in section IV. In
	section IV, we discover many complex YM instanton solutions with
	rank$\beta=0,1$ (sheaves) and $2$ (bundle) respectively. These solutions are
	all of $\det=0$, and thus are new and not considered previously. For
	illustration, we give examples of explicit complex YM $2$-instanton solutions
	for each case. In addition, an example of new extended complex YM
	$2$-instanton field strength for the sheaf case with rank$\beta=0$ and
	$\det=0$ will be given in Eq.(\ref{v5}). Finally, the conclusion will be given
	in section V. \ \ \ \ \ \ \ \ \ \ \ \ \ \ 
	
	\section{Review of YM $2$-instanton sheaves with rank$\beta=1$ and $\det\neq
		0$}
	
	In this section, we briefly review the biquaternion construction of $SU(2)$
	complex ADHM instantons \cite{Ann,Ann2}. We will use three approaches to
	construct the complex YM instantons. We will pay attention to the existence of
	sheaf structures of YM $2$-instanton sheaves \cite{Ann,Ann2,Ann3} in the monad construction.
	
	\subsection{The Biquaternion construction of $SU(2)$ complex ADHM instantons}
	
	In this approach, the first step was to introduce the $(k+1)\times k$
	biquarternion matrix $\Delta(x)=a+bx$%
	
	\begin{equation}
	\Delta(x)_{ab}=a_{ab}+b_{ab}x,\text{ }a_{ab}=a_{ab}^{\mu}e_{\mu},b_{ab}%
	=b_{ab}^{\mu}e_{\mu} \label{ab}%
	\end{equation}
	where $a_{ab}^{\mu}$ and $b_{ab}^{\mu}$ are complex numbers, and $a_{ab}$ and
	$b_{ab}$ are biquarternions. In the biquaternion construction of $SU(2)$
	complex ADHM instantons, the quadratic condition on the biquaternion matrix
	$\Delta(x)$ of $SU(2)$ complex instantons reads%
	
	\begin{equation}
	\Delta(x)^{\circledast}\Delta(x)=f^{-1}=\text{symmetric, non-singular }k\times
	k\text{ matrix for }x\notin J \label{con}%
	\end{equation}
	where for $x\in J,$%
	\begin{equation}
	\det\Delta(x)^{\circledast}\Delta(x)=0. \label{points}%
	\end{equation}
	The set $J$ is called jumping lines. An explicit example of jumping lines will
	be given in Eq.(\ref{zero}). The biconjugation in Eq.(\ref{points}) is defined
	in Eq.(\ref{bi}) below. There are no jumping lines for $SU(2)$ real instantons
	on $S^{4}$.
	
	The biconjugation \cite{Ham} of a biquaternion%
	\begin{equation}
	z=z_{\mu}e_{\mu}\text{, \ }z_{\mu}\in C,
	\end{equation}
	is defined to be%
	\begin{equation}
	z^{\circledast}=z_{\mu}e_{\mu}^{\dagger}=z_{0}e_{0}-z_{1}e_{1}-z_{2}%
	e_{2}-z_{3}e_{3}=x^{\dagger}+y^{\dagger}i. \label{bi}%
	\end{equation}
	In some occasion, the unit quarternions can be expressed as Pauli matrices%
	\begin{equation}
	e_{0}\rightarrow%
	\begin{pmatrix}
	1 & 0\\
	0 & 1
	\end{pmatrix}
	,e_{i}\rightarrow-i\sigma_{i}\ \text{; }i=1,2,3.
	\end{equation}
	We will use the norm square of a biquarternion, and it is defined to be%
	\begin{equation}
	|z|_{c}^{2}=z^{\circledast}z=(z_{0})^{2}+(z_{1})^{2}+(z_{2})^{2}+(z_{3})^{2},
	\end{equation}
	which is a \textit{complex} number in general. This property will turn out to
	be important for the construction of the new extended complex YM instantons in
	this paper. See Eq.(\ref{poss}) in section III.
	
	It was shown that \cite{Ann2} there are no sheaf structures for the $SU(2)$
	complex YM $1$-instanton neither complex diagonal $k$-instantons (for $k=2$
	case with $\det\neq0$). So, for simplicity, we will choose $k=2$ to calculate
	and identify complex YM instanton sheaf structures of the $SU(2)$ (diagonal
	and non-diagonal) complex YM $2$-instantons with $\det=0$ in this paper.
	
	\subsection{The $SU(2)$ complex ADHM equations}
	
	The second method to construct $SU(2)$ complex ADHM data is to solve the
	complex ADHM equations \cite{Donald,5}%
	\begin{subequations}
		\begin{align}
		\left[  B_{11},B_{12}\right]  +I_{1}J_{1}  &  =0,\label{adhm1}\\
		\left[  B_{21},B_{22}\right]  +I_{2}J_{2}  &  =0,\label{adhm2}\\
		\left[  B_{11},B_{22}\right]  +\left[  B_{21},B_{12}\right]  +I_{1}J_{2}%
		+I_{2}J_{1}  &  =0. \label{adhm3}%
		\end{align}
		In this approach, one key step is to use the explicit matrix representation
		(EMR) \cite{Ann2} of the biquaternion and do the rearrangement rule
		\cite{Ann2} to \bigskip explicitly identify the complex ADHM data
		$(B_{lm},I_{m,}J_{m})$ with $l,m=1,2$ from the $\Delta(x)$ matrix in
		Eq.(\ref{ab}).
		
		For illustration, we calculate the $SU(2)$ diagonal $2$-instanton case. In the
		EMR, a biquaternion can be written as a $2\times2$ complex matrix
	\end{subequations}
	\begin{align}
	z  &  =z^{0}e_{0}+z^{1}e_{1}+z^{2}e_{2}+z^{3}e_{3}\nonumber\\
	&  =%
	\begin{bmatrix}
	\left(  a^{0}+b^{3}\right)  +i\left(  b^{0}-a^{3}\right)  & \left(
	-a^{2}+b^{1}\right)  +i\left(  -b^{2}-a^{1}\right) \\
	\left(  a^{2}+b^{1}\right)  +i\left(  b^{2}-a^{1}\right)  & \left(
	a^{0}-b^{3}\right)  +i\left(  b^{0}+a^{3}\right)
	\end{bmatrix}
	\end{align}
	where $a^{\mu}$ and $b^{\mu}$ are real and imaginary parts of $z^{\mu}$
	respectively. For the diagonal $2$-instanton%
	
	\begin{align}
	a  &  =%
	\begin{bmatrix}
	\lambda_{1} & \lambda_{2}\\
	y_{11} & 0\\
	0 & y_{22}%
	\end{bmatrix}
	=%
	\begin{bmatrix}
	p_{1}+iq_{1} & 0 & p_{2}+iq_{2} & 0\\
	0 & p_{1}+iq_{1} & 0 & p_{2}+iq_{2}\\
	y_{11}^{0}-iy_{11}^{3} & -\left(  y_{11}^{2}+iy_{11}^{1}\right)  & 0 & 0\\
	y_{11}^{2}-iy_{11}^{1} & y_{11}^{0}+iy_{11}^{3} & 0 & 0\\
	0 & 0 & y_{22}^{0}-iy_{22}^{3} & -\left(  y_{22}^{2}+iy_{22}^{1}\right) \\
	0 & 0 & y_{22}^{2}-iy_{22}^{1} & y_{22}^{0}+iy_{22}^{3}%
	\end{bmatrix}
	\\
	&  \rightarrow%
	\begin{bmatrix}
	p_{1}+iq_{1} & p_{2}+iq_{2} & 0 & 0\\
	0 & 0 & p_{1}+iq_{1} & p_{2}+iq_{2}\\
	y_{11}^{0}-iy_{11}^{3} & 0 & -\left(  y_{11}^{2}+iy_{11}^{1}\right)  & 0\\
	0 & y_{22}^{0}-iy_{22}^{3} & 0 & -\left(  y_{22}^{2}+iy_{22}^{1}\right) \\
	y_{11}^{2}-iy_{11}^{1} & 0 & y_{11}^{0}+iy_{11}^{3} & 0\\
	0 & y_{22}^{2}-iy_{22}^{1} & 0 & y_{22}^{0}+iy_{22}^{3}%
	\end{bmatrix}
	=%
	\begin{bmatrix}
	J_{1} & J_{2}\\
	B_{11} & B_{21}\\
	B_{12} & B_{22}%
	\end{bmatrix}
	\label{2nd}%
	\end{align}
	where in Eq.(\ref{2nd}) we have performed the \textit{rearrangement rule }for
	an element $z_{ij}$ in $a$\textit{ }
	\begin{align}
	z_{2n-1,2m-1}  &  \rightarrow z_{n,m}\text{ },\nonumber\\
	z_{2n-1,2m}  &  \rightarrow z_{n,k+m}\text{ },\nonumber\\
	z_{2n,2m-1}  &  \rightarrow z_{k+n,m}\text{ },\nonumber\\
	z_{2n,2m}  &  \rightarrow z_{k+n,k+m}.
	\end{align}
	The EMR and the rearrangement rule for $a^{\circledast}$ can be similarly
	performed. With the identification in Eq.(\ref{2nd}) (and a similar one for
	$a^{\circledast}$), one can show that the quadratic condition in
	Eq.(\ref{con}) leads to the complex ADHM equations in Eq.(\ref{adhm1}) to
	Eq.(\ref{adhm3}).
	
	For the $SU(2)$ real ADHM instantons, one imposes the conditions%
	\begin{subequations}
		\begin{align}
		I_{1}  &  =J^{\dagger},I_{2}=-I,J_{1}=I^{\dagger},J_{2}=J,\nonumber\\
		B_{11}  &  =B_{2}^{\dagger},B_{12}=B_{1}^{\dagger},B_{21}=-B_{1},B_{22}=B_{2}%
		\end{align}
		to recover the real ADHM equations%
	\end{subequations}
	\begin{subequations}
		\begin{align}
		\left[  B_{1},B_{2}\right]  +IJ  &  =0,\\
		\left[  B_{1},B_{1}^{\dagger}\right]  +\left[  B_{2},B_{2}^{\dagger}\right]
		+II^{\dagger}-J^{\dagger}J  &  =0.
		\end{align}

		\subsection{The monad construction and YM $2$-instanton sheaves}
		
		The third method to construct $SU(2)$ complex ADHM instantons is the monad
		construction. This method is particular suitable for constructing instanton
		sheaves. One introduces the $\bigskip\alpha\,$and $\beta$ matrices as
		functions of homogeneous coordinates $z,w,x,y$ of $CP^{3}$ and defines%
	\end{subequations}
	\begin{subequations}
		\begin{align}
		\alpha &  =%
		\begin{bmatrix}
		zB_{11}+wB_{21}+x\\
		zB_{12}+wB_{22}+y\\
		zJ_{1}+wJ_{2}%
		\end{bmatrix}
		,\label{alpha}\\
		\beta &  =%
		\begin{bmatrix}
		-zB_{12}-wB_{22}-y & zB_{11}+wB_{21}+x & zI_{1}+wI_{2}%
		\end{bmatrix}
		. \label{beta}%
		\end{align}
		It can be shown that the condition%
	\end{subequations}
	\begin{equation}
	\beta\alpha=0
	\end{equation}
	is satisfied if and only if the complex ADHM equations in Eq.(\ref{adhm1}) to
	Eq.(\ref{adhm3}) holds.
	
	In the monad construction of the holomorphic vector bundles, either $\beta$ is
	not surjective or $\alpha$ is not injective at some points of $CP^{3}$ for
	some ADHM data, the dimension of $($Ker $\beta$/ Im $\alpha)$ varies from
	point to point on $CP^{3}$, and one encounters "instanton sheaves" on $CP^{3}$
	\cite{math2}. In our previous publication \cite{Ann2}, we discovered that for
	some ADHM data at some sheaf points on $CP^{3}$, there exists a common
	eigenvector $u$ in the costable condition $\alpha u=0$ or \cite{math2}%
	\begin{subequations}
		\begin{align}
		\left(  zB_{11}+wB_{21}\right)  u  &  =-xu,\label{a}\\
		\left(  zB_{12}+wB_{22}\right)  u  &  =-yu,\label{b}\\
		\left(  zJ_{1}+wJ_{2}\right)  u  &  =0. \label{c}%
		\end{align}
		So $\alpha$ is not injective there and the dimension of $($Ker $\beta$/ Im
		$\alpha)$ is not a constant over $CP^{3}$. Similar discussion can be done for
		cases with $\beta$ not surjective \cite{Ann4,math2}. That is, for some ADHM
		data at some sheaf points on $CP^{3}$, there exists a common eigenvector $v$
		in the stable condition \cite{math2}%
	\end{subequations}
	\begin{equation}
	v\beta=0. \label{d}%
	\end{equation}
	We will choose to work on rank$\beta$ in this paper.
	
	The first example of YM instanton sheaf discovered in \cite{Ann2} was the
	$2$-instanton sheaf. For points $[x:y:z:w]=[0:0:1:1]$ on $CP^{3}$ with the
	ADHM data\bigskip%
	\begin{equation}%
	\begin{bmatrix}
	\lambda_{1} & \lambda_{2}\\
	y_{11} & y_{12}\\
	y_{12} & y_{22}%
	\end{bmatrix}
	=%
	\begin{bmatrix}
	a & 0 & 0 & ia\\
	0 & a & ia & 0\\
	\frac{-i}{\sqrt{2}}a & 0 & 0 & \frac{a}{\sqrt{2}}\\
	0 & \frac{-i}{\sqrt{2}}a & \frac{a}{\sqrt{2}} & 0\\
	0 & \frac{a}{\sqrt{2}} & \frac{i}{\sqrt{2}}a & 0\\
	\frac{a}{\sqrt{2}} & 0 & 0 & \frac{i}{\sqrt{2}}a
	\end{bmatrix}
	,a\in C,a\neq0, \label{data1}%
	\end{equation}
	it was shown that $\alpha$ is not injective.
	
	Moreover, to understand the change of dimensionality of vector bundles at the
	sheaf points, we can calculate $\alpha$ and $\beta$ at the sheaf point
	$[x:y:z:w]=[0:0:1:1]$ to be $(a\neq0)$%
	\begin{equation}
	\alpha_{\lbrack0:0:1:1]}=%
	\begin{pmatrix}
	\frac{-ia}{\sqrt{2}} & \frac{a}{\sqrt{2}}\\
	\frac{a}{\sqrt{2}} & \frac{ia}{\sqrt{2}}\\
	\frac{-ia}{\sqrt{2}} & \frac{a}{\sqrt{2}}\\
	\frac{a}{\sqrt{2}} & \frac{ia}{\sqrt{2}}\\
	a & ia\\
	a & ia
	\end{pmatrix}
	,\beta_{\lbrack0:0:1:1]}=%
	\begin{pmatrix}
	\frac{ia}{\sqrt{2}} & -\frac{a}{\sqrt{2}} & \frac{-ia}{\sqrt{2}} & \frac
	{a}{\sqrt{2}} & -a & a\\
	-\frac{a}{\sqrt{2}} & \frac{-ia}{\sqrt{2}} & \frac{a}{\sqrt{2}} & \frac
	{ia}{\sqrt{2}} & -ia & ia
	\end{pmatrix}
	, \label{beta2}%
	\end{equation}
	which are both of rank $1$. So the dimensions of Im$\alpha_{\lbrack0:0:1:1]}$
	and Ker$\beta_{\lbrack0:0:1:1]}$ are $1$ and $5$ (or rank$\beta=1$)
	respectively, which imply the dimension of the quotient space%
	\begin{equation}
	\dim(\text{Ker}\beta_{\lbrack0:0:1:1]}/\operatorname{Im}\alpha_{\lbrack
		0:0:1:1]})=5-1=4. \label{5}%
	\end{equation}
	Note that for points other than sheaf points, $\dim($Ker$\beta
	/\operatorname{Im}\alpha)=4-2=2$.
	
	\section{Non-existence of YM $2$-instanton sheaves with rank$\beta=0$ and
		$\det\neq0$}
	
	\bigskip Following the discovery of YM $2$-instanton sheaves with
	rank$\beta=1$ on some points of $CP^{3}$ \cite{Ann2} (see Eq.(\ref{beta2})),
	it is very natural to look for the rest case of YM $2$-instanton sheaves with
	rank$\beta=0$ on some points of $CP^{3}$ with given ADHM data. We first define
	the ADHM data as%
	\begin{align}
	y_{11}  &  \equiv%
	\begin{pmatrix}
	d_{1} & d_{2}\\
	d_{3} & d_{4}%
	\end{pmatrix}
	,y_{22}=-%
	\begin{pmatrix}
	a_{1} & a_{2}\\
	a_{3} & a_{4}%
	\end{pmatrix}
	,\\
	\lambda_{1}  &  =%
	\begin{pmatrix}
	\lambda_{1}^{0}-i\lambda_{1}^{3} & -\left(  \lambda_{1}^{2}+i\lambda_{1}%
	^{1}\right) \\
	\lambda_{1}^{2}-i\lambda_{1}^{1} & \lambda_{1}^{0}+i\lambda_{1}^{3}%
	\end{pmatrix}
	,\lambda_{1}^{\circledast}=%
	\begin{pmatrix}
	\lambda_{1}^{0}+i\lambda_{1}^{3} & \lambda_{1}^{2}+i\lambda_{1}^{1}\\
	-\left(  \lambda_{1}^{2}-i\lambda_{1}^{1}\right)  & \lambda_{1}^{0}%
	-i\lambda_{1}^{3}%
	\end{pmatrix}
	,\\
	\lambda_{2}  &  =%
	\begin{pmatrix}
	\lambda_{2}^{0}-i\lambda_{2}^{3} & -\left(  \lambda_{2}^{2}+i\lambda_{2}%
	^{1}\right) \\
	\lambda_{2}^{2}-i\lambda_{2}^{1} & \lambda_{2}^{0}+i\lambda_{2}^{3}%
	\end{pmatrix}
	,\lambda_{2}^{\circledast}=%
	\begin{pmatrix}
	\lambda_{2}^{0}+i\lambda_{2}^{3} & \lambda_{2}^{2}+i\lambda_{2}^{1}\\
	-\left(  \lambda_{2}^{2}-i\lambda_{2}^{1}\right)  & \lambda_{2}^{0}%
	-i\lambda_{2}^{3}%
	\end{pmatrix}
	.
	\end{align}
	Eq.(\ref{con}) then implies \cite{Ann,CSW}%
	\begin{equation}
	y_{12}=y_{21}=\frac{1}{2}\frac{\left(  y_{11}-y_{22}\right)  }{\left\vert
		y_{11}-y_{22}\right\vert _{c}^{2}}\left(  \lambda_{2}^{\circledast}\lambda
	_{1}-\lambda_{1}^{\circledast}\lambda_{2}\right)  . \label{csw}%
	\end{equation}

	The denominator of $y_{12}$ can be calculated to be%
	\begin{align}
	y_{11}-y_{22}  &  =%
	\begin{pmatrix}
	d_{1}+a_{1} & d_{2}+a_{2}\\
	d_{3}+a_{3} & d_{4}+a_{4}%
	\end{pmatrix}
	,\nonumber\\
	\left\vert y_{11}-y_{22}\right\vert _{c}^{2}  &  =\left(  d_{1}+a_{1}\right)
	\left(  d_{4}+a_{4}\right)  -\left(  d_{2}+a_{2}\right)  \left(  d_{3}%
	+a_{3}\right) \nonumber\\
	&  =\det. \label{aft}%
	\end{align}
	The vanishing of the $\det$ defined above will be used to search for the new
	YM $2$-instanton solutions in section IV.
	
	It is important to note that since in general $\left\vert y_{11}%
	-y_{22}\right\vert _{c}^{2}$ is a \textit{complex} number, the vanishing of
	$\det$ defined above does not mean $y_{11}=y_{22}$. While in the real
	instanton case \cite{CSW}, $\left\vert y_{11}-y_{22}\right\vert ^{2}$ is a
	non-negative real number and $\left\vert y_{11}-y_{22}\right\vert ^{2}=0$
	implies $y_{11}=y_{22}$ which is not allowed for the $2$-instanton solutions.
	So the extended $SU(2)$ complex YM instantons discovered in this paper have no
	real instanton counterparts.
	
	To simplify the calculation, one introduces
	\begin{equation}
	l=%
	\begin{vmatrix}
	\lambda_{1}^{0} & \lambda_{1}^{3}\\
	\lambda_{2}^{0} & \lambda_{2}^{3}%
	\end{vmatrix}
	-%
	\begin{vmatrix}
	\lambda_{1}^{1} & \lambda_{1}^{2}\\
	\lambda_{2}^{1} & \lambda_{2}^{2}%
	\end{vmatrix}
	,m=%
	\begin{vmatrix}
	\lambda_{1}^{0} & \lambda_{1}^{1}\\
	\lambda_{2}^{0} & \lambda_{2}^{1}%
	\end{vmatrix}
	-%
	\begin{vmatrix}
	\lambda_{1}^{2} & \lambda_{1}^{3}\\
	\lambda_{2}^{2} & \lambda_{2}^{3}%
	\end{vmatrix}
	,n=%
	\begin{vmatrix}
	\lambda_{1}^{0} & \lambda_{1}^{2}\\
	\lambda_{2}^{0} & \lambda_{2}^{2}%
	\end{vmatrix}
	-%
	\begin{vmatrix}
	\lambda_{1}^{3} & \lambda_{1}^{1}\\
	\lambda_{2}^{3} & \lambda_{2}^{1}%
	\end{vmatrix}
	, \label{lmn}%
	\end{equation}
	which imply%
	\begin{equation}
	\left(  \lambda_{2}^{\circledast}\lambda_{1}-\lambda_{1}^{\circledast}%
	\lambda_{2}\right)  =2i%
	\begin{pmatrix}
	l & m-in\\
	m+in & -l
	\end{pmatrix}
	.
	\end{equation}
	Finally we end up with the expression%
	\begin{align}
	y_{12}  &  =\frac{1}{2}\frac{\left(  y_{11}-y_{22}\right)  }{\left\vert
		y_{11}-y_{22}\right\vert _{c}^{2}}\left(  \lambda_{2}^{\circledast}\lambda
	_{1}-\lambda_{1}^{\circledast}\lambda_{2}\right) \nonumber\\
	&  =\frac{i}{\det}%
	\begin{bmatrix}
	\left(  d_{1}+a_{1}\right)  l+\left(  d_{2}+a_{2}\right)  \left(  m+in\right)
	& \left(  d_{1}+a_{1}\right)  \left(  m-in\right)  -\left(  d_{2}%
	+a_{2}\right)  l\\
	\left(  d_{3}+a_{3}\right)  l+\left(  d_{4}+a_{4}\right)  \left(  m+in\right)
	& \left(  d_{3}+a_{3}\right)  \left(  m-in\right)  -\left(  d_{4}%
	+a_{4}\right)  l
	\end{bmatrix}
	.
	\end{align}

	We are now ready to identify the ADHM data. The EMR of the $y$ data is%
	\begin{equation}%
	\begin{bmatrix}%
	\begin{pmatrix}
	d_{1} & d_{2}\\
	d_{3} & d_{4}%
	\end{pmatrix}
	& \frac{i}{\det}%
	\begin{bmatrix}%
	\begin{array}
	{c}%
	\left(  d_{1}+a_{1}\right)  l\\
	+\left(  d_{2}+a_{2}\right)  \left(  m+in\right)
	\end{array}
	&
	\begin{array}
	{c}%
	\left(  d_{1}+a_{1}\right)  \left(  m-in\right) \\
	-\left(  d_{2}+a_{2}\right)  l
	\end{array}
	\\%
	\begin{array}
	{c}%
	\left(  d_{3}+a_{3}\right)  l\\
	+\left(  d_{4}+a_{4}\right)  \left(  m+in\right)
	\end{array}
	&
	\begin{array}
	{c}%
	\left(  d_{3}+a_{3}\right)  \left(  m-in\right) \\
	-\left(  d_{4}+a_{4}\right)  l
	\end{array}
	\end{bmatrix}
	\\
	\frac{i}{\det}%
	\begin{bmatrix}%
	\begin{array}
	{c}%
	\left(  d_{1}+a_{1}\right)  l\\
	+\left(  d_{2}+a_{2}\right)  \left(  m+in\right)
	\end{array}
	&
	\begin{array}
	{c}%
	\left(  d_{1}+a_{1}\right)  \left(  m-in\right) \\
	-\left(  d_{2}+a_{2}\right)  l
	\end{array}
	\\%
	\begin{array}
	{c}%
	\left(  d_{3}+a_{3}\right)  l\\
	+\left(  d_{4}+a_{4}\right)  \left(  m+in\right)
	\end{array}
	&
	\begin{array}
	{c}%
	\left(  d_{3}+a_{3}\right)  \left(  m-in\right) \\
	-\left(  d_{4}+a_{4}\right)  l
	\end{array}
	\end{bmatrix}
	& -%
	\begin{pmatrix}
	a_{1} & a_{2}\\
	a_{3} & a_{4}%
	\end{pmatrix}
	\end{bmatrix}
	.
	\end{equation}
	After imposing the rearrangement rule, we can identify the following ADHM data%
	
	\begin{align}
	B_{11}  &  =%
	\begin{pmatrix}
	d_{1} & \frac{i}{\det}\left[
	\begin{array}
	{c}%
	\left(  d_{1}+a_{1}\right)  l\\
	+\left(  d_{2}+a_{2}\right)  \left(  m+in\right)
	\end{array}
	\right] \\
	\frac{i}{\det}\left[
	\begin{array}
	{c}%
	\left(  d_{1}+a_{1}\right)  l\\
	+\left(  d_{2}+a_{2}\right)  \left(  m+in\right)
	\end{array}
	\right]  & -a_{1}%
	\end{pmatrix}
	,\\
	B_{21}  &  =%
	\begin{pmatrix}
	d_{2} & \frac{i}{\det}\left[
	\begin{array}
	{c}%
	\left(  d_{1}+a_{1}\right)  \left(  m-in\right) \\
	-\left(  d_{2}+a_{2}\right)  l
	\end{array}
	\right] \\
	\frac{i}{\det}\left[
	\begin{array}
	{c}%
	\left(  d_{1}+a_{1}\right)  \left(  m-in\right) \\
	-\left(  d_{2}+a_{2}\right)  l
	\end{array}
	\right]  & -a_{2}%
	\end{pmatrix}
	,\\
	B_{12}  &  =%
	\begin{pmatrix}
	d_{3} & \frac{i}{\det}\left[
	\begin{array}
	{c}%
	\left(  d_{3}+a_{3}\right)  l\\
	+\left(  d_{4}+a_{4}\right)  \left(  m+in\right)
	\end{array}
	\right] \\
	\frac{i}{\det}\left[
	\begin{array}
	{c}%
	\left(  d_{3}+a_{3}\right)  l\\
	+\left(  d_{4}+a_{4}\right)  \left(  m+in\right)
	\end{array}
	\right]  & -a_{3}%
	\end{pmatrix}
	,\\
	B_{22}  &  =%
	\begin{pmatrix}
	d_{4} & \frac{i}{\det}\left[
	\begin{array}
	{c}%
	\left(  d_{3}+a_{3}\right)  \left(  m-in\right) \\
	-\left(  d_{4}+a_{4}\right)  l
	\end{array}
	\right] \\
	\frac{i}{\det}\left[
	\begin{array}
	{c}%
	\left(  d_{3}+a_{3}\right)  \left(  m-in\right) \\
	-\left(  d_{4}+a_{4}\right)  l
	\end{array}
	\right]  & -a_{4}%
	\end{pmatrix}
	.
	\end{align}
	and%
	\begin{align}
	I_{1}  &  =%
	\begin{pmatrix}
	-\lambda_{1}^{2}+i\lambda_{1}^{1} & \lambda_{1}^{0}-i\lambda_{1}^{3}\\
	-\lambda_{2}^{2}+i\lambda_{2}^{1} & \lambda_{2}^{0}-i\lambda_{2}^{3}%
	\end{pmatrix}
	,\\
	I_{2}  &  =%
	\begin{pmatrix}
	-\left(  \lambda_{1}^{0}+i\lambda_{1}^{3}\right)  & -\left(  \lambda_{1}%
	^{2}+i\lambda_{1}^{1}\right) \\
	-\left(  \lambda_{2}^{0}+i\lambda_{2}^{3}\right)  & -\left(  \lambda_{2}%
	^{2}+i\lambda_{2}^{1}\right)
	\end{pmatrix}
	.
	\end{align}

	We can now consider the rank$\beta=0$ or $\beta=0$ case.
	
	(1) For those points on $CP^{3}$ with $z=1$, $\beta=0$ means%
	
	\begin{align}
	B_{12}+wB_{22}+y  &  =0,\label{b1}\\
	B_{11}+wB_{21}+x  &  =0,\label{b2}\\
	I_{1}+wI_{2}  &  =0. \label{ii}%
	\end{align}
	By using Eq.(\ref{b1}) and Eq.(\ref{b2}), we obtain%
	\begin{equation}%
	\begin{pmatrix}
	d_{3}+wd_{4}+y & \frac{i}{\det}\left\{
	\begin{array}
	{c}%
	\left[  \left(  d_{3}+a_{3}\right)  l+\left(  d_{4}+a_{4}\right)  \left(
	m+in\right)  \right] \\
	+w\left[  \left(  d_{3}+a_{3}\right)  \left(  m-in\right)  -\left(
	d_{4}+a_{4}\right)  l\right]
	\end{array}
	\right\} \\
	\frac{i}{\det}\left\{
	\begin{array}
	{c}%
	\left[  \left(  d_{3}+a_{3}\right)  l+\left(  d_{4}+a_{4}\right)  \left(
	m+in\right)  \right] \\
	+w\left[  \left(  d_{3}+a_{3}\right)  \left(  m-in\right)  -\left(
	d_{4}+a_{4}\right)  l\right]
	\end{array}
	\right\}  & -a_{3}-wa_{4}+y
	\end{pmatrix}
	=0. \label{g1}%
	\end{equation}
	and%
	\begin{equation}%
	\begin{pmatrix}
	d_{1}+wd_{2}+x & \frac{i}{\det}\left\{
	\begin{array}
	{c}%
	\left[  \left(  d_{1}+a_{3}\right)  l+\left(  d_{2}+a_{2}\right)  \left(
	m+in\right)  \right] \\
	+w\left[  \left(  d_{1}+a_{3}\right)  \left(  m-in\right)  -\left(
	d_{2}+a_{2}\right)  l\right]
	\end{array}
	\right\} \\
	\frac{i}{\det}\left\{
	\begin{array}
	{c}%
	\left[  \left(  d_{1}+a_{3}\right)  l+\left(  d_{2}+a_{2}\right)  \left(
	m+in\right)  \right] \\
	+w\left[  \left(  d_{1}+a_{3}\right)  \left(  m-in\right)  -\left(
	d_{2}+a_{2}\right)  l\right]
	\end{array}
	\right\}  & -a_{1}-wa_{2}+x
	\end{pmatrix}
	=0. \label{g2}%
	\end{equation}
	respectively. We note that Eq.(\ref{g1}) gives%
	\begin{equation}
	d_{3}+wd_{4}+y=-a_{3}-wa_{4}+y=0,
	\end{equation}
	or%
	\begin{equation}
	w=\frac{d_{3}+a_{3}}{-a_{4}-d_{4}}. \label{w1}%
	\end{equation}
	Eq.(\ref{g2}) gives%
	\begin{equation}
	d_{1}+wd_{2}+x=-a_{3}-wa_{4}+x=0,
	\end{equation}
	or%
	\begin{equation}
	w=\frac{-\left(  a_{1}+d_{1}\right)  }{d_{2}+a_{2}}. \label{w2}%
	\end{equation}
	We see from Eq.(\ref{w1}) and Eq.(\ref{w2}) that%
	\begin{equation}
	\frac{d_{3}+a_{3}}{-a_{4}-d_{4}}=\frac{-\left(  a_{1}+d_{1}\right)  }%
	{d_{2}+a_{2}},
	\end{equation}
	which means%
	\begin{equation}
	\det=0.
	\end{equation}

	We will call instanton solutions with $\det=0$ the "extended complex YM
	instanton solutions" which were not considered in the previous construction
	\cite{Ann}. Moreover, as already mentioned in the paragraph after
	Eq.(\ref{aft}), since in general $\left\vert y_{11}-y_{22}\right\vert _{c}%
	^{2}$ is a \textit{complex} number, the vanishing of $\det$ does not mean
	$y_{11}=y_{22}$, or%
	\begin{equation}
	\det=\left\vert y_{11}-y_{22}\right\vert _{c}^{2}=0\nRightarrow y_{11}=y_{22}.
	\label{poss}%
	\end{equation}
	So it is possible to have complex YM $2$-instanton solutions with $\det=0$.
	While in the previous $SU(2)$ real instanton case \cite{CSW}, $\left\vert
	y_{11}-y_{22}\right\vert ^{2}$ is a non-negative real number and%
	\begin{equation}
	\left\vert y_{11}-y_{22}\right\vert ^{2}=0\Rightarrow y_{11}=y_{22}
	\label{poss2}%
	\end{equation}
	which is not allowed for the real YM $2$-instanton solutions. So the extended
	$SU(2)$ complex YM instantons discovered in this paper have no real instanton counterparts.
	
	\bigskip(2) For those points on $CP^{3}$ with $z=0,w=1,$ $\beta=0$ means%
	
	\begin{align}
	-B_{22}-y  &  =0,\label{b0}\\
	B_{21}+x  &  =0,\label{b00}\\
	I_{2}  &  =0. \label{i0}%
	\end{align}
	We see that Eq.(\ref{b0}) gives
	\begin{equation}%
	\begin{bmatrix}
	d_{4}+y & \frac{i}{\det}\left[  \left(  d_{3}+a_{3}\right)  \left(
	m-in\right)  -\left(  d_{4}+a_{4}\right)  l\right] \\
	\frac{i}{\det}\left[  \left(  d_{3}+a_{3}\right)  \left(  m-in\right)
	-\left(  d_{4}+a_{4}\right)  l\right]  & -\beta+y
	\end{bmatrix}
	=0,
	\end{equation}
	which implies%
	\begin{equation}
	d_{4}=-a_{4}=-y. \label{k1}%
	\end{equation}

	On the other hand, Eq.(\ref{b00}) gives%
	\begin{equation}%
	\begin{bmatrix}
	d_{2}+x & \frac{i}{\det}\left[  \left(  d_{1}+a_{3}\right)  \left(
	m-in\right)  -\left(  d_{2}+a_{2}\right)  l\right] \\
	\frac{i}{\det}\left[  \left(  d_{1}+a_{3}\right)  \left(  m-in\right)
	-\left(  d_{2}+a_{2}\right)  l\right]  & -a_{2}+x
	\end{bmatrix}
	=0,
	\end{equation}
	which implies
	\begin{equation}
	d_{2}=-a_{2}=-x. \label{k2}%
	\end{equation}
	Eq.(\ref{k1}) and Eq.(\ref{k2}) imply%
	\begin{equation}
	\det=\left(  d_{1}+a_{1}\right)  \left(  d_{4}+a_{4}\right)  -\left(
	d_{2}+a_{2}\right)  \left(  d_{3}+a_{3}\right)  =0,
	\end{equation}
	which again corresponds to the extended case.
	
	(3) Finally for $z=0$ and $w=0$, $\beta=0$ in Eq.(\ref{beta}) gives $x=0=y.$
	Thus for this case there is no point on $CP^{3}$ which supports rank$\beta=0$.
	
	We thus have completed the proof that rank$\beta=0$ or $\beta=0$ implies
	$\det=0$. So there is no rank$\beta=0$ with $\det\neq0$ complex YM instanton
	sheaf solutions. In all our previous construction of complex YM instantons
	(sheaves) \cite{Ann,Ann2} , we have assumed $\det\neq0$ as in the real YM
	instanton case. In the next section, we are looking for $\det=0$ (see
	Eq.(\ref{poss})) complex YM instanton solutions both for the sheaf
	(rank$\beta=0,1$ on some points of $CP^{3}$) and bundle (rank$\beta=2$ on the
	whole $CP^{3}$) cases.
	
	\section{Extended YM $2$-instanton solutions with $\det=0$}
	
	In this section, we are looking for complex YM $2$-instanton solutions with%
	\begin{equation}
	\det=\left\vert y_{11}-y_{22}\right\vert _{c}^{2}=\left(  d_{1}+a_{1}\right)
	\left(  d_{4}+a_{4}\right)  -\left(  d_{2}+a_{2}\right)  d_{3}+a_{3})=0.
	\end{equation}
	The idea is that whether one can factor out a $\det$ factor from the nemurator
	of the rhs of Eq.(\ref{csw}). These solutions were not considered in
	\cite{Ann}, and if exist, they correspond to "limiting solutions" living on
	the boundary of the moduli space. We will discuss the solutions with three
	different cases with rank$\beta=0,1$ on some points of $CP^{3}$ with some
	given ADHM data (sheaves) and rank$\beta=2$ on the whole $CP^{3}$ for some
	other given ADHM data (bundle) respectively.
	
	\subsection{Sheaf case with rank$\beta=0$}
	
	For the first case, we consider $\beta=0$%
	\begin{align}
	zB_{12}+wB_{22}+y  &  =0,\label{b11}\\
	zB_{11}+wB_{21}+x  &  =0,\label{b22}\\
	zI_{1}+wI_{2}  &  =0. \label{iii}%
	\end{align}
	and assuming $z\neq0$, $w\neq0$ on $CP^{3}$. We first note that Eq.(\ref{iii})
	or%
	\begin{equation}
	I_{1}+\frac{w}{z}I_{2}=0
	\end{equation}
	can be used to solve half of the number of $\lambda$ parameters. Indeed, the
	vanishing of the upper two components of%
	\begin{equation}%
	\begin{pmatrix}
	-\lambda_{1}^{2}+i\lambda_{1}^{1}-\frac{w}{z}\left(  \lambda_{1}^{0}%
	+i\lambda_{1}^{3}\right)  & \lambda_{1}^{0}-i\lambda_{1}^{3}-\frac{w}%
	{z}\left(  \lambda_{1}^{2}+i\lambda_{1}^{1}\right) \\
	-\lambda_{2}^{2}+i\lambda_{2}^{1}-\frac{w}{z}\left(  \lambda_{2}^{0}%
	+i\lambda_{2}^{3}\right)  & \lambda_{2}^{0}-i\lambda_{2}^{3}-\frac{w}%
	{z}\left(  \lambda_{2}^{2}+i\lambda_{2}^{1}\right)
	\end{pmatrix}
	=0
	\end{equation}
	give%
	\begin{align}
	-\lambda_{1}^{2}+i\lambda_{1}^{1}  &  =\frac{w}{z}\lambda_{1}^{0}+i\frac{w}%
	{z}\lambda_{1}^{3},\\
	\lambda_{1}^{2}+i\lambda_{1}^{1}  &  =(\frac{w}{z})^{-1}\lambda_{1}%
	^{0}-i(\frac{w}{z})^{-1}\lambda_{1}^{3},
	\end{align}
	which can be used to get%
	\begin{align}
	\lambda_{1}^{2}  &  =\frac{1}{2}\left[  \left(  (\frac{w}{z})^{-1}-\frac{w}%
	{z}\right)  \lambda_{1}^{0}-i\left(  (\frac{w}{z})^{-1}+\frac{w}{z}\right)
	\lambda_{1}^{3}\right] \nonumber\\
	&  =\frac{1}{2}\left(  \delta\lambda_{1}^{0}-i\rho\lambda_{1}^{3}\right)
	\end{align}
	and
	\begin{align}
	\lambda_{1}^{1}  &  =\frac{1}{2i}\left[  \left(  (\frac{w}{z})^{-1}+\frac
	{w}{z}\right)  \lambda_{1}^{0}-i\left(  (\frac{w}{z})^{-1}-\frac{w}{z}\right)
	\lambda_{1}^{3}\right] \nonumber\\
	&  =\frac{1}{2i}\left(  \rho\lambda_{1}^{0}-i\delta\lambda_{1}^{3}\right)
	\end{align}
	where we have defined%
	\begin{equation}
	\delta=(\frac{w}{z})^{-1}-\frac{w}{z},\rho=(\frac{w}{z})^{-1}+\frac{w}{z}.
	\end{equation}

	Similarly the vanishing of the lower two components of Eq.(\ref{ii}) solves
	another two $\lambda$, and we end up with%
	\begin{align}
	\lambda_{1}^{2}  &  =\frac{1}{2}\left(  \delta\lambda_{1}^{0}-i\rho\lambda
	_{1}^{3}\right)  ,\label{la1}\\
	\lambda_{1}^{1}  &  =\frac{1}{2i}\left(  \rho\lambda_{1}^{0}-i\delta
	\lambda_{1}^{3}\right)  ,\label{la2}\\
	\lambda_{2}^{2}  &  =\frac{1}{2}\left(  \delta\lambda_{2}^{0}-i\rho\lambda
	_{2}^{3}\right)  ,\label{la3}\\
	\lambda_{2}^{1}  &  =\frac{1}{2i}\left(  \rho\lambda_{2}^{0}-i\delta
	\lambda_{2}^{3}\right)  . \label{la4}%
	\end{align}
	With the expressions in Eq.(\ref{la1}) to Eq.(\ref{la4}), one can simplify
	$l,m,n$ to obtain%
	\begin{align}
	l  &  =2%
	\begin{vmatrix}
	\lambda_{1}^{0} & \lambda_{1}^{3}\\
	\lambda_{2}^{0} & \lambda_{2}^{3}%
	\end{vmatrix}
	,\\
	m  &  =-\delta%
	\begin{vmatrix}
	\lambda_{1}^{0} & \lambda_{1}^{3}\\
	\lambda_{2}^{0} & \lambda_{2}^{3}%
	\end{vmatrix}
	,\\
	n  &  =-i\rho%
	\begin{vmatrix}
	\lambda_{1}^{0} & \lambda_{1}^{3}\\
	\lambda_{2}^{0} & \lambda_{2}^{3}%
	\end{vmatrix}
	.
	\end{align}

	One can now easily calculate the expression%
	\begin{equation}
	\lambda_{2}^{\circledast}\lambda_{1}-\lambda_{1}^{\circledast}\lambda_{2}=4i%
	\begin{vmatrix}
	\lambda_{1}^{0} & \lambda_{1}^{3}\\
	\lambda_{2}^{0} & \lambda_{2}^{3}%
	\end{vmatrix}%
	\begin{bmatrix}
	1 & -(\frac{w}{z})^{-1}\\
	\frac{w}{z} & -1
	\end{bmatrix}
	\end{equation}
	to obtain%
	\begin{align}
	y_{12}  &  =\frac{1}{2}\frac{\left(  y_{11}-y_{22}\right)  }{\left\vert
		y_{11}-y_{22}\right\vert _{c}^{2}}\left(  \lambda_{2}^{\circledast}\lambda
	_{1}-\lambda_{1}^{\circledast}\lambda_{2}\right) \\
	&  =\frac{2i%
		\begin{vmatrix}
		\lambda_{1}^{0} & \lambda_{1}^{3}\\
		\lambda_{2}^{0} & \lambda_{2}^{3}%
		\end{vmatrix}
	}{\left\vert y_{11}-y_{22}\right\vert _{c}^{2}}%
	\begin{pmatrix}
	d_{1}+a_{1} & d_{2}+a_{2}\\
	d_{3}+a_{3} & d_{4}+a_{4}%
	\end{pmatrix}%
	\begin{pmatrix}
	1 & -(\frac{w}{z})^{-1}\\
	\frac{w}{z} & -1
	\end{pmatrix}
	\label{csw2}\\
	&  =\frac{2i%
		\begin{vmatrix}
		\lambda_{1}^{0} & \lambda_{1}^{3}\\
		\lambda_{2}^{0} & \lambda_{2}^{3}%
		\end{vmatrix}
	}{\left\vert y_{11}-y_{22}\right\vert _{c}^{2}}%
	\begin{pmatrix}
	\left(  d_{1}+a_{1}\right)  +\frac{w}{z}\left(  d_{2}+a_{2}\right)  &
	-(\frac{w}{z})^{-1}\left(  d_{1}+a_{1}\right)  -\left(  d_{2}+a_{2}\right) \\
	\left(  d_{3}+a_{3}\right)  +\frac{w}{z}\left(  d_{4}+a_{4}\right)  &
	-(\frac{w}{z})^{-1}\left(  d_{3}+a_{3}\right)  -\left(  d_{4}+a_{4}\right)
	\end{pmatrix}
	. \label{csw3}%
	\end{align}
	Note that the first column of the matrix in Eq.(\ref{csw3}) is proportional to
	the second column with proportional constant $-(\frac{w}{z})^{-1}$ due to the
	structure of the second matrix in Eq.(\ref{csw2}).
	
	To continue the calculation, we begin with general value of $\det$, and after
	factoring out a $\det$ factor from the nemurator of the rhs of Eq.(\ref{csw2}%
	), we will take the $\det\rightarrow0$ limit. There are two channels to
	achieve this factorization\bigskip.
	
	\subsubsection{Factorization I}
	
	We first note that the $\det$ can be written as
	\begin{equation}
	\det=\left(  d_{4}+a_{4}\right)  \left[  \left(  d_{1}+a_{1}\right)
	-\frac{\left(  d_{2}+a_{2}\right)  \left(  d_{3}+a_{3}\right)  }{\left(
		d_{4}+a_{4}\right)  }\right]
	\end{equation}
	or%
	\begin{equation}
	\left(  \frac{1}{d_{4}+a_{4}}\right)  \det=\left[  \left(  d_{1}+a_{1}\right)
	-\left(  d_{2}+a_{2}\right)  \frac{\left(  d_{3}+a_{3}\right)  }{\left(
		d_{4}+a_{4}\right)  }\right]  . \label{non}%
	\end{equation}

	On the other hand, we remember that%
	\begin{align*}
	zB_{12}+wB_{22}+y  &  =0,\\
	zB_{11}+wB_{21}+x  &  =0
	\end{align*}
	imply%
	\begin{equation}
	\frac{w}{z}=\frac{d_{3}+a_{3}}{-a_{4}-d_{4}}=\frac{-\left(  a_{1}%
		+d_{1}\right)  }{d_{2}+a_{2}}, \label{res}%
	\end{equation}
	which means $\det=0$. Let's now consider nonvanishing $\det$ (with $\beta
	\neq0$) and rewrite Eq.(\ref{non}) as
	\begin{equation}
	\left(  \frac{1}{d_{4}+a_{4}}\right)  \det=\left[  \left(  d_{1}+a_{1}\right)
	+\left(  d_{2}+a_{2}\right)  \left(  \frac{w}{z}+\varepsilon^{\prime}\right)
	\right]  .
	\end{equation}
	Note that as $\beta\rightarrow0$, we have $\det\rightarrow0,$ which gives%
	\begin{equation}
	\frac{w}{z}=-\frac{\left(  d_{3}+a_{3}\right)  }{\left(  d_{4}+a_{4}\right)
	}=-\frac{\left(  a_{1}+d_{1}\right)  }{d_{2}+a_{2}} \label{wz}%
	\end{equation}
	and%
	\begin{equation}
	\varepsilon^{\prime}\rightarrow0.
	\end{equation}
	We can rewrite $\left(  d_{2}+a_{2}\right)  \varepsilon^{\prime}%
	=\varepsilon\det$ with $\varepsilon$ a finite number as $\det\rightarrow0,$
	and obtain
	\begin{equation}
	\left(  \frac{1}{d_{4}+a_{4}}+\varepsilon\right)  \det=\left[  \left(
	d_{1}+a_{1}\right)  +\frac{w}{z}\left(  d_{2}+a_{2}\right)  \right]  .
	\label{sub}%
	\end{equation}

	We can now use Eq.(\ref{sub}) to express $\frac{w}{z}$ in terms of $\det$ as%
	\begin{equation}
	\frac{w}{z}=\frac{-\left(  d_{3}+a_{3}\right)  }{d_{4}+a_{4}}+\frac{\left(
		\varepsilon\cdot\det\right)  }{d_{2}+a_{2}},
	\end{equation}
	and calculate%
	\begin{equation}
	\left(  d_{3}+a_{3}\right)  +\frac{w}{z}\left(  d_{4}+a_{4}\right)  =\left(
	\frac{\varepsilon\det\left(  d_{4}+a_{4}\right)  }{d_{2}+a_{2}}\right)  .
	\end{equation}

	We conclude that Eq.(\ref{csw3}) can be written as%
	\begin{align}
	y_{12}  &  =\frac{2i}{\det}%
	\begin{pmatrix}
	\left(  \varepsilon+\frac{1}{d_{4}+a_{4}}\right)  \det & -(\frac{w}{z}%
	)^{-1}\left(  \varepsilon+\frac{1}{d_{4}+a_{4}}\right)  \det\\
	\left(  \frac{\varepsilon\left(  d_{4}+a_{4}\right)  }{d_{2}+a_{2}}\right)
	\det & -(\frac{w}{z})^{-1}\left(  \frac{\varepsilon\left(  d_{4}+a_{4}\right)
	}{d_{2}+a_{2}}\right)  \det
	\end{pmatrix}%
	\begin{vmatrix}
	\lambda_{1}^{0} & \lambda_{1}^{3}\\
	\lambda_{2}^{0} & \lambda_{2}^{3}%
	\end{vmatrix}
	\label{dd}\\
	&  =%
	\begin{pmatrix}
	c_{1} & c_{2}\\
	c_{3} & c_{4}%
	\end{pmatrix}
	\label{cc}%
	\end{align}
	with%
	\begin{align}
	c_{1}  &  =2i\left(  \varepsilon+\frac{1}{d_{4}+a_{4}}\right)
	\begin{vmatrix}
	\lambda_{1}^{0} & \lambda_{1}^{3}\\
	\lambda_{2}^{0} & \lambda_{2}^{3}%
	\end{vmatrix}
	,\\
	c_{3}  &  =2i\left(  \frac{\varepsilon\left(  d_{4}+a_{4}\right)  }%
	{d_{2}+a_{2}}\right)
	\begin{vmatrix}
	\lambda_{1}^{0} & \lambda_{1}^{3}\\
	\lambda_{2}^{0} & \lambda_{2}^{3}%
	\end{vmatrix}
	,\\%
	\begin{pmatrix}
	c_{2}\\
	c_{4}%
	\end{pmatrix}
	&  =-(\frac{w}{z})^{-1}%
	\begin{pmatrix}
	c_{1}\\
	c_{3}%
	\end{pmatrix}
	. \label{c3}%
	\end{align}
	It is important to note that to achieve the factorization, $\frac{w}{z}$ is an
	arbitrary number (except $z\neq0$, $w\neq0$ on $CP^{3}$) and $\varepsilon$ is
	finite as $\det\rightarrow0$ in Eq.(\ref{dd}).
	
	To obtain Eq.(\ref{cc}), we have cancelled out the $\det$ factor and take the
	$\det\rightarrow0$ limit. We thus have explicitly shown the existence of the
	extended complex YM $2$-instanton sheaf solutions with $\det=0$ and
	rank$\beta=0$. These new complex YM instanton solutions were not considered in
	\cite{Ann}, and live on the boundary of the moduli space. Their existences
	strongly depend on the structure of the matrix in Eq.(\ref{csw2}).
	
	Finally we need to check the validities of Eq.(\ref{b11}) and Eq.(\ref{b22}).
	The $B$ matrice of the ADHM data can be written as
	\begin{align}
	B_{11}  &  =%
	\begin{pmatrix}
	d_{1} & c_{1}\\
	c_{1} & -a_{1}%
	\end{pmatrix}
	,B_{21}=%
	\begin{pmatrix}
	d_{2} & c_{2}\\
	c_{2} & -a_{2}%
	\end{pmatrix}
	,\\
	B_{12}  &  =%
	\begin{pmatrix}
	d_{3} & c_{3}\\
	c_{3} & -a_{3}%
	\end{pmatrix}
	,B_{22}=%
	\begin{pmatrix}
	d_{4} & c_{4}\\
	c_{4} & -a_{4}%
	\end{pmatrix}
	.
	\end{align}
	The vanishing of the diagonal terms of%
	\begin{equation}
	zB_{11}+wB_{21}+x=%
	\begin{pmatrix}
	d_{1}z+d_{2}w+x & c_{1}z+c_{2}w\\
	c_{1}z+c_{2}w & -a_{1}z-a_{2}w+x
	\end{pmatrix}
	=0 \label{c1}%
	\end{equation}
	and%
	\begin{equation}
	zB_{12}+wB_{22}+y=%
	\begin{pmatrix}
	d_{3}z+d_{4}w+y & c_{3}z+c_{4}w\\
	c_{3}z+c_{4}w & -a_{3}z-a_{4}w+y
	\end{pmatrix}
	=0 \label{c2}%
	\end{equation}
	mean Eq.(\ref{wz}), which implies $\det=0$, and%
	\begin{align}
	x  &  =-d_{1}z-wd_{2},\\
	y  &  =-d_{3}z-wd_{4}.
	\end{align}
	The vanishing of the off-diagonal terms of Eq.(\ref{c1}) and Eq.(\ref{c2})
	mean%
	\begin{align}
	c_{1}+\frac{w}{z}c_{2}  &  =0,\\
	c_{3}+\frac{w}{z}c_{4}  &  =0,
	\end{align}
	which are results of Eq.(\ref{c3}). Thus the solution set for this case
	contain $11$ parameters. $7=4+4-1$ from $a_{j}$ and $d_{j}$ subject to
	$\det=0$, and $4$ more parameters from $\lambda_{1}^{0},\lambda_{1}%
	^{3},\lambda_{2}^{0}$ and $\lambda_{2}^{3}.$
	
	For completeness and to include the cases of $z=0$ or $w=0,$ we have
	explicitly checked the validity of the complex ADHM equations of the above
	ADHM data.
	
	In the end of this section, we give one explicit example of the extended YM
	$2$-instanton sheaf for this case. Let's begin with the following ADHM data%
	\begin{equation}
	y_{11}=%
	\begin{pmatrix}
	d_{1} & d_{2}\\
	d_{3} & d_{4}%
	\end{pmatrix}
	=%
	\begin{pmatrix}
	1 & 0\\
	0 & 1
	\end{pmatrix}
	,y_{22}=-%
	\begin{pmatrix}
	a_{1} & a_{2}\\
	a_{3} & a_{4}%
	\end{pmatrix}
	=%
	\begin{pmatrix}
	0 & -p\\
	-p & 0
	\end{pmatrix}
	,
	\end{equation}
	which gives%
	\begin{equation}
	\det=1-p^{2}.
	\end{equation}
	For half of the $\lambda$ ADHM data, we assume%
	\begin{equation}
	\lambda_{1}^{0}=1,\lambda_{1}^{3}=1,\lambda_{2}^{0}=i,\lambda_{2}^{3}=-i.
	\label{lam1}%
	\end{equation}
	With the expressions in Eq.(\ref{la1}) to Eq.(\ref{la4}) and the following
	information
	\begin{equation}
	\frac{w}{z}=-\left(  \frac{d_{3}+a_{3}}{d_{4}+a_{4}}\right)  \rightarrow
	1,\frac{z}{w}\rightarrow1\text{ as }p\rightarrow-1,
	\end{equation}
	the other half of $\lambda$ ADHM data can be calculated to be%
	\begin{equation}
	\lambda_{1}^{1}=-i,\lambda_{1}^{2}=-i,\lambda_{2}^{1}=1,\lambda_{2}^{2}=-1.
	\label{lam2}%
	\end{equation}

	We can now use the following data%
	\begin{align}
	\lambda_{1}  &  =%
	\begin{pmatrix}
	\lambda_{1}^{0}-i\lambda_{1}^{3} & -\left(  \lambda_{1}^{2}+i\lambda_{1}%
	^{1}\right) \\
	\lambda_{1}^{2}-i\lambda_{1}^{1} & \lambda_{1}^{0}+i\lambda_{1}^{3}%
	\end{pmatrix}
	=%
	\begin{pmatrix}
	1-i & -1+i\\
	-1-i & 1+i
	\end{pmatrix}
	,\nonumber\\
	\lambda_{1}^{\circledast}  &  =%
	\begin{pmatrix}
	\lambda_{1}^{0}+i\lambda_{1}^{3} & \lambda_{1}^{2}+i\lambda_{1}^{1}\\
	-\left(  \lambda_{1}^{2}-i\lambda_{1}^{1}\right)  & \lambda_{1}^{0}%
	-i\lambda_{1}^{3}%
	\end{pmatrix}
	=%
	\begin{pmatrix}
	1+i & 1-i\\
	1+i & 1-i
	\end{pmatrix}
	,\nonumber\\
	\lambda_{2}  &  =%
	\begin{pmatrix}
	\lambda_{2}^{0}-i\lambda_{2}^{3} & -\left(  \lambda_{2}^{2}+i\lambda_{2}%
	^{1}\right) \\
	\lambda_{2}^{2}-i\lambda_{2}^{1} & \lambda_{2}^{0}+i\lambda_{2}^{3}%
	\end{pmatrix}
	=%
	\begin{pmatrix}
	i-1 & 1-i\\
	-1-i & i+1
	\end{pmatrix}
	,\nonumber\\
	\lambda_{2}^{\circledast}  &  =%
	\begin{pmatrix}
	\lambda_{2}^{0}+i\lambda_{2}^{3} & \lambda_{2}^{2}+i\lambda_{2}^{1}\\
	-\left(  \lambda_{2}^{2}-i\lambda_{2}^{1}\right)  & \lambda_{2}^{0}%
	-i\lambda_{2}^{3}%
	\end{pmatrix}
	=%
	\begin{pmatrix}
	1+i & -1+i\\
	1+i & -1+i
	\end{pmatrix}
	\end{align}
	to calculate $y_{12}$ and obtain%
	\begin{align}
	y_{12}  &  =\frac{1}{2}\frac{\left(  y_{11}-y_{22}\right)  }{\left\vert
		y_{11}-y_{22}\right\vert _{c}^{2}}\left(  \lambda_{2}^{\circledast}\lambda
	_{1}-\lambda_{1}^{\circledast}\lambda_{2}\right) \nonumber\\
	&  =4\frac{%
		\begin{pmatrix}
		1 & p\\
		p & 1
		\end{pmatrix}
	}{1-p^{2}}%
	\begin{pmatrix}
	1 & -1\\
	1 & -1
	\end{pmatrix}
	=\frac{4}{1-p^{2}}%
	\begin{pmatrix}
	\frac{1-p^{2}}{1-p} & -\left(  \frac{1-p^{2}}{1-p}\right) \\
	\frac{1-p^{2}}{1-p} & -\left(  \frac{1-p^{2}}{1-p}\right)
	\end{pmatrix}
	\nonumber\\
	&  =%
	\begin{pmatrix}
	\frac{4}{1-p} & -\left(  \frac{4}{1-p}\right) \\
	\frac{4}{1-p} & -\left(  \frac{4}{1-p}\right)
	\end{pmatrix}
	=%
	\begin{pmatrix}
	c_{1} & c_{2}\\
	c_{3} & c_{4}%
	\end{pmatrix}
	\end{align}
	where we have seen the factorization of the $\det=1-p^{2}$ factor as expected.
	
	In the $p\rightarrow-1$ limit, we can calculate the following $y$ matrices%
	\begin{equation}
	y_{11}=%
	\begin{pmatrix}
	d_{1} & d_{2}\\
	d_{3} & d_{4}%
	\end{pmatrix}
	=%
	\begin{pmatrix}
	1 & 0\\
	0 & 1
	\end{pmatrix}
	,y_{22}=-%
	\begin{pmatrix}
	a_{1} & a_{2}\\
	a_{3} & a_{4}%
	\end{pmatrix}
	=%
	\begin{pmatrix}
	0 & 1\\
	1 & 0
	\end{pmatrix}
	,y_{12}=%
	\begin{pmatrix}
	c_{1} & c_{2}\\
	c_{3} & c_{4}%
	\end{pmatrix}
	=%
	\begin{pmatrix}
	2 & -2\\
	2 & -2
	\end{pmatrix}
	. \label{yy}%
	\end{equation}
	One can now use Eq.(\ref{lam1}), Eq.(\ref{lam2}) and Eq.(\ref{yy}) to
	calculate the following ADHM data
	\begin{align}
	J_{1}  &  =%
	\begin{pmatrix}
	\lambda_{1}^{0}-i\lambda_{1}^{3} & \lambda_{2}^{0}-i\lambda_{2}^{3}\\
	\lambda_{1}^{2}-i\lambda_{1}^{1} & \lambda_{2}^{2}-i\lambda_{2}^{1}%
	\end{pmatrix}
	=%
	\begin{pmatrix}
	1-i & i-1\\
	-1-i & -1-i
	\end{pmatrix}
	,\nonumber\\
	J_{2}  &  =%
	\begin{pmatrix}
	-\left(  \lambda_{1}^{2}+i\lambda_{1}^{1}\right)  & -\left(  \lambda_{2}%
	^{2}+i\lambda_{2}^{1}\right) \\
	\lambda_{1}^{0}+i\lambda_{1}^{3} & \lambda_{2}^{0}+i\lambda_{2}^{3}%
	\end{pmatrix}
	=%
	\begin{pmatrix}
	-1+i & 1-i\\
	1+i & i+1
	\end{pmatrix}
	,\nonumber\\
	I_{1}  &  =%
	\begin{pmatrix}
	-\left(  \lambda_{1}^{2}-i\lambda_{1}^{1}\right)  & \lambda_{1}^{0}%
	-i\lambda_{1}^{3}\\
	-\left(  \lambda_{2}^{2}-i\lambda_{2}^{1}\right)  & \lambda_{2}^{0}%
	-i\lambda_{2}^{3}%
	\end{pmatrix}
	=%
	\begin{pmatrix}
	1+i & 1-i\\
	1+i & -1+i
	\end{pmatrix}
	,\nonumber\\
	I_{2}  &  =%
	\begin{pmatrix}
	-\left(  \lambda_{1}^{0}+i\lambda_{1}^{3}\right)  & -\left(  \lambda_{1}%
	^{2}+i\lambda_{1}^{1}\right) \\
	-\left(  \lambda_{2}^{0}+i\lambda_{2}^{3}\right)  & -\left(  \lambda_{2}%
	^{2}+i\lambda_{2}^{1}\right)
	\end{pmatrix}
	=%
	\begin{pmatrix}
	-1-i & -1+i\\
	-1-i & 1-i
	\end{pmatrix}
	,
	\end{align}
	and%
	\begin{align}
	B_{11}  &  =%
	\begin{pmatrix}
	d_{1} & c_{1}\\
	c_{1} & -a_{1}%
	\end{pmatrix}
	=%
	\begin{pmatrix}
	1 & 2\\
	2 & 0
	\end{pmatrix}
	,B_{21}=%
	\begin{pmatrix}
	d_{2} & c_{2}\\
	c_{2} & -a_{2}%
	\end{pmatrix}
	=%
	\begin{pmatrix}
	0 & -2\\
	-2 & 1
	\end{pmatrix}
	,\nonumber\\
	B_{12}  &  =%
	\begin{pmatrix}
	d_{3} & c_{3}\\
	c_{3} & -a_{3}%
	\end{pmatrix}
	=%
	\begin{pmatrix}
	0 & 2\\
	2 & 1
	\end{pmatrix}
	,B_{22}=%
	\begin{pmatrix}
	d_{4} & c_{4}\\
	c_{4} & -a_{4}%
	\end{pmatrix}
	=%
	\begin{pmatrix}
	1 & -2\\
	-2 & 0
	\end{pmatrix}
	.
	\end{align}

	We can now check the three complex ADHM equations%
	\begin{align}
	&  \left[  B_{11},B_{12}\right]  +I_{1}J_{1}\nonumber\\
	&  =%
	\begin{pmatrix}
	1 & 2\\
	2 & 0
	\end{pmatrix}%
	\begin{pmatrix}
	0 & 2\\
	2 & 1
	\end{pmatrix}
	-%
	\begin{pmatrix}
	0 & 2\\
	2 & 1
	\end{pmatrix}%
	\begin{pmatrix}
	1 & 2\\
	2 & 0
	\end{pmatrix}
	+%
	\begin{pmatrix}
	1+i & 1-i\\
	1+i & -1+i
	\end{pmatrix}%
	\begin{pmatrix}
	1-i & i-1\\
	-1-i & -1-i
	\end{pmatrix}
	\nonumber\\
	&  =%
	\begin{pmatrix}
	4 & 4\\
	0 & 4
	\end{pmatrix}
	-%
	\begin{pmatrix}
	4 & 0\\
	4 & 4
	\end{pmatrix}
	+%
	\begin{pmatrix}
	0 & -4\\
	4 & 0
	\end{pmatrix}
	=%
	\begin{pmatrix}
	0 & 0\\
	0 & 0
	\end{pmatrix}
	,
	\end{align}%
	\begin{align}
	&  \left[  B_{21},B_{22}\right]  +I_{2}J_{2}\nonumber\\
	&  =%
	\begin{pmatrix}
	0 & -2\\
	-2 & 1
	\end{pmatrix}%
	\begin{pmatrix}
	1 & -2\\
	-2 & 0
	\end{pmatrix}
	-%
	\begin{pmatrix}
	1 & -2\\
	-2 & 0
	\end{pmatrix}%
	\begin{pmatrix}
	0 & -2\\
	-2 & 1
	\end{pmatrix}
	+%
	\begin{pmatrix}
	-1-i & -1+i\\
	-1-i & 1-i
	\end{pmatrix}%
	\begin{pmatrix}
	-1+i & 1-i\\
	1+i & i+1
	\end{pmatrix}
	\\
	&  =%
	\begin{pmatrix}
	4 & 0\\
	-4 & 4
	\end{pmatrix}
	-%
	\begin{pmatrix}
	4 & -4\\
	0 & 4
	\end{pmatrix}
	+%
	\begin{pmatrix}
	0 & -4\\
	4 & 0
	\end{pmatrix}
	=%
	\begin{pmatrix}
	0 & 0\\
	0 & 0
	\end{pmatrix}
	\nonumber
	\end{align}
	and%
	\begin{align}
	&  [B_{11},B_{22}]+\left[  B_{21},B_{12}\right]  +I_{1}.J_{2}+I_{2}%
	J_{1}\nonumber\\
	&  =%
	\begin{pmatrix}
	1 & 2\\
	2 & 0
	\end{pmatrix}%
	\begin{pmatrix}
	1 & -2\\
	-2 & 0
	\end{pmatrix}
	-%
	\begin{pmatrix}
	1 & -2\\
	-2 & 0
	\end{pmatrix}%
	\begin{pmatrix}
	1 & 2\\
	2 & 0
	\end{pmatrix}
	+%
	\begin{pmatrix}
	0 & -2\\
	-2 & 1
	\end{pmatrix}%
	\begin{pmatrix}
	0 & 2\\
	2 & 1
	\end{pmatrix}
	-%
	\begin{pmatrix}
	0 & 2\\
	2 & 1
	\end{pmatrix}%
	\begin{pmatrix}
	0 & -2\\
	-2 & 1
	\end{pmatrix}
	\nonumber\\
	&  +%
	\begin{pmatrix}
	1+i & 1-i\\
	1+i & -1+i
	\end{pmatrix}%
	\begin{pmatrix}
	-1+i & 1-i\\
	1+i & i+1
	\end{pmatrix}
	+%
	\begin{pmatrix}
	-1-i & -1+i\\
	-1-i & 1-i
	\end{pmatrix}%
	\begin{pmatrix}
	1-i & i-1\\
	-1-i & -1-i
	\end{pmatrix}
	\nonumber\\
	&  =%
	\begin{pmatrix}
	-3 & -2\\
	2 & -4
	\end{pmatrix}
	-%
	\begin{pmatrix}
	-3 & 2\\
	-2 & -4
	\end{pmatrix}
	+%
	\begin{pmatrix}
	-4 & -2\\
	2 & -3
	\end{pmatrix}
	-%
	\begin{pmatrix}
	-4 & 2\\
	-2 & -3
	\end{pmatrix}
	+%
	\begin{pmatrix}
	0 & 4\\
	-4 & 0
	\end{pmatrix}
	+%
	\begin{pmatrix}
	0 & 4\\
	-4 & 0
	\end{pmatrix}
	\nonumber\\
	&  =%
	\begin{pmatrix}
	0 & 0\\
	0 & 0
	\end{pmatrix}
	.
	\end{align}

	It remains to check%
	
	\begin{align}
	zB_{11}+wB_{21}+x  &  =%
	\begin{pmatrix}
	z+x & 2z-2w\\
	2z-2w & w+x
	\end{pmatrix}
	=0,\nonumber\\
	zB_{12}+wB_{22}+y  &  =%
	\begin{pmatrix}
	w+y & 2z-2w\\
	2z-2w & z+y
	\end{pmatrix}
	=0. \label{sol}%
	\end{align}
	The solutions of Eq.(\ref{sol}) are
	\begin{equation}
	w=z,x=-z,y=-z,
	\end{equation}
	which represent a point on $CP^{3}$%
	\begin{equation}
	\lbrack x:y:z:w]=[-1:-1:1:1].
	\end{equation}
	This is the point where jumping occurs.
	
	\subsubsection{Factorization II}
	
	For the second case, we first assume $\det\neq0$ and set%
	\begin{equation}%
	\begin{vmatrix}
	\lambda_{1}^{0} & \lambda_{1}^{3}\\
	\lambda_{2}^{0} & \lambda_{2}^{3}%
	\end{vmatrix}
	=\det\cdot\tilde{\varepsilon}.
	\end{equation}
	After cancelling out the $\det$ factor and take the $\det\rightarrow0$ limit,
	Eq.(\ref{csw3}) can then be written as%
	
	\begin{align}
	y_{12}  &  =\frac{2i%
		\begin{vmatrix}
		\lambda_{1}^{0} & \lambda_{1}^{3}\\
		\lambda_{2}^{0} & \lambda_{2}^{3}%
		\end{vmatrix}
	}{\left\vert y_{11}-y_{22}\right\vert _{c}^{2}}%
	\begin{pmatrix}
	d_{1}+a_{1} & d_{2}+a_{2}\\
	d_{3}+a_{3} & d_{4}+a_{4}%
	\end{pmatrix}%
	\begin{pmatrix}
	1 & -(\frac{w}{z})^{-1}\\
	\frac{w}{z} & -1
	\end{pmatrix}
	\nonumber\\
	&  =2i\tilde{\varepsilon}%
	\begin{pmatrix}
	\left(  d_{1}+a_{1}\right)  +\frac{w}{z}\left(  d_{2}+a_{2}\right)  &
	-(\frac{w}{z})^{-1}\left(  d_{1}+a_{1}\right)  -\left(  d_{2}+a_{2}\right) \\
	\left(  d_{3}+a_{3}\right)  +\frac{w}{z}\left(  d_{4}+a_{4}\right)  &
	-(\frac{w}{z})^{-1}\left(  d_{3}+a_{3}\right)  -\left(  d_{4}+a_{4}\right)
	\end{pmatrix}
	\label{no}\\
	&  =%
	\begin{pmatrix}
	0 & 0\\
	0 & 0
	\end{pmatrix}
	\label{000}%
	\end{align}
	where we have used Eq.(\ref{wz}) in the $\det\rightarrow0$ limit. The result
	of Eq.(\ref{000}) is not surprising since in the calculation of case I, we
	already knew that in the $\det\rightarrow0$ limit, the matrix in Eq.(\ref{no})
	factored out a $\det$ factor. However, it is surprising to see that for the
	extended complex YM $2$-instantons with diagonal $y$ ADHM data, there exist YM
	instanton sheaf structure! For the $\det\neq0$ complex YM instantons
	constructed previously \cite{Ann}, it was shown \cite{Ann2} that there is no
	instanton sheaf solutions with diagonal $y$ ADHM data.
	
	As in the case I, we need to check the validities of Eq.(\ref{b11}) and
	Eq.(\ref{b22}). The vanishing of the diagonal terms of%
	\begin{align}
	zB_{11}+wB_{21}+x  &  =0,\label{c11}\\
	zB_{12}+wB_{22}+y  &  =0 \label{c22}%
	\end{align}
	mean Eq.(\ref{wz}) which implies $\det=0$, and%
	\begin{align}
	x  &  =-d_{1}z-wd_{2},\\
	y  &  =-d_{3}z-wd_{4}.
	\end{align}
	On the other hand, the vanishing of the off-diagonal terms of Eq.(\ref{c11})
	and Eq.(\ref{c22}) are results of Eq.(\ref{c3}). There is another constraint
	from the following equation%
	\begin{equation}%
	\begin{vmatrix}
	\lambda_{1}^{0} & \lambda_{1}^{3}\\
	\lambda_{2}^{0} & \lambda_{2}^{3}%
	\end{vmatrix}
	=0.
	\end{equation}

	In the following we give one explicit example of YM $2$-instanton sheaf for
	this case. Similar to the example for the case I, let's take the following
	ADHM data%
	\begin{equation}
	y_{11}=%
	\begin{pmatrix}
	d_{1} & d_{2}\\
	d_{3} & d_{4}%
	\end{pmatrix}
	=%
	\begin{pmatrix}
	1 & 0\\
	0 & 1
	\end{pmatrix}
	,y_{22}=-%
	\begin{pmatrix}
	a_{1} & a_{2}\\
	a_{3} & a_{4}%
	\end{pmatrix}
	=%
	\begin{pmatrix}
	0 & -p\\
	-p & 0
	\end{pmatrix}
	, \label{dat1}%
	\end{equation}
	which gives%
	\begin{equation}
	\det=1-p^{2}.
	\end{equation}
	We assume (for simplicity, take $\tilde{\varepsilon}=1$)%
	\begin{equation}%
	\begin{vmatrix}
	\lambda_{1}^{0} & \lambda_{1}^{3}\\
	\lambda_{2}^{0} & \lambda_{2}^{3}%
	\end{vmatrix}
	=%
	\begin{vmatrix}
	1 & p\\
	p & 1
	\end{vmatrix}
	=1-p^{2}=\det\cdot\tilde{\varepsilon}. \label{dat2}%
	\end{equation}
	Since%
	\begin{equation}
	\frac{w}{z}=-\left(  \frac{d_{3}+a_{3}}{d_{4}+a_{4}}\right)  =-p\rightarrow
	1,\frac{z}{w}\rightarrow1\text{ as }p\rightarrow-1,
	\end{equation}
	we have%
	\begin{equation}
	\delta=(\frac{w}{z})^{-1}-\frac{w}{z}\rightarrow0,\rho=(\frac{w}{z}%
	)^{-1}+\frac{w}{z}\rightarrow2.
	\end{equation}
	So all other parameters can be calculated to be%
	\begin{align}
	\lambda_{1}^{2}  &  =\frac{1}{2}\left(  \delta\lambda_{1}^{0}-i\rho\lambda
	_{1}^{3}\right)  =i,\nonumber\\
	\lambda_{1}^{1}  &  =\frac{1}{2i}\left(  \rho\lambda_{1}^{0}-i\delta
	\lambda_{1}^{3}\right)  =-i,\nonumber\\
	\lambda_{2}^{2}  &  =\frac{1}{2}\left(  \delta\lambda_{2}^{0}-i\rho\lambda
	_{2}^{3}\right)  =-i,\nonumber\\
	\lambda_{2}^{1}  &  =\frac{1}{2i}\left(  \rho\lambda_{2}^{0}-i\delta
	\lambda_{2}^{3}\right)  =i \label{dat3}%
	\end{align}
	as $p\rightarrow-1$. Finally $y_{12}$ can be calculated to be%
	\begin{align}
	y_{12}  &  =\frac{1}{2}\frac{\left(  y_{11}-y_{22}\right)  }{\left\vert
		y_{11}-y_{22}\right\vert _{c}^{2}}\left(  \lambda_{2}^{\circledast}\lambda
	_{1}-\lambda_{1}^{\circledast}\lambda_{2}\right) \nonumber\\
	&  =\frac{1}{2}\frac{%
		\begin{pmatrix}
		1 & p\\
		p & 1
		\end{pmatrix}
	}{1-a^{2}}%
	\begin{pmatrix}
	1 & -1\\
	1 & -1
	\end{pmatrix}%
	\begin{vmatrix}
	\lambda_{1}^{0} & \lambda_{1}^{3}\\
	\lambda_{2}^{0} & \lambda_{2}^{3}%
	\end{vmatrix}
	=\frac{1}{2}\frac{%
		\begin{pmatrix}
		1 & p\\
		p & 1
		\end{pmatrix}
	}{1-p^{2}}%
	\begin{pmatrix}
	1 & -1\\
	1 & -1
	\end{pmatrix}
	(1-p^{2})\tilde{\varepsilon}\nonumber\\
	&  =\frac{1}{2}%
	\begin{pmatrix}
	1 & -1\\
	-1 & 1
	\end{pmatrix}%
	\begin{pmatrix}
	1 & -1\\
	1 & -1
	\end{pmatrix}
	\tilde{\varepsilon}=%
	\begin{pmatrix}
	0 & 0\\
	0 & 0
	\end{pmatrix}
	\label{dat4}%
	\end{align}
	as expected.
	
	The next step is to calculate ADHM data. One easily obtains%
	\begin{align}
	J_{1}  &  =%
	\begin{pmatrix}
	\lambda_{1}^{0}-i\lambda_{1}^{3} & \lambda_{2}^{0}-i\lambda_{2}^{3}\\
	\lambda_{1}^{2}-i\lambda_{1}^{1} & \lambda_{2}^{2}-i\lambda_{2}^{1}%
	\end{pmatrix}
	=%
	\begin{pmatrix}
	1+i & -i-1\\
	-1+i & 1-i
	\end{pmatrix}
	,\nonumber\\
	J_{2}  &  =%
	\begin{pmatrix}
	-\left(  \lambda_{1}^{2}+i\lambda_{1}^{1}\right)  & -\left(  \lambda_{2}%
	^{2}+i\lambda_{2}^{1}\right) \\
	\lambda_{1}^{0}+i\lambda_{1}^{3} & \lambda_{2}^{0}+i\lambda_{2}^{3}%
	\end{pmatrix}
	=%
	\begin{pmatrix}
	-1-i & 1+i\\
	1-i & i-1
	\end{pmatrix}
	=-J_{1},\nonumber\\
	I_{1}  &  =%
	\begin{pmatrix}
	-\left(  \lambda_{1}^{2}-i\lambda_{1}^{1}\right)  & \lambda_{1}^{0}%
	-i\lambda_{1}^{3}\\
	-\left(  \lambda_{2}^{2}-i\lambda_{2}^{1}\right)  & \lambda_{2}^{0}%
	-i\lambda_{2}^{3}%
	\end{pmatrix}
	=%
	\begin{pmatrix}
	1-i & 1+i\\
	-1+i & -1-i
	\end{pmatrix}
	,\nonumber\\
	I_{2}  &  =%
	\begin{pmatrix}
	-\left(  \lambda_{1}^{0}+i\lambda_{1}^{3}\right)  & -\left(  \lambda_{1}%
	^{2}+i\lambda_{1}^{1}\right) \\
	-\left(  \lambda_{2}^{0}+i\lambda_{2}^{3}\right)  & -\left(  \lambda_{2}%
	^{2}+i\lambda_{2}^{1}\right)
	\end{pmatrix}
	=%
	\begin{pmatrix}
	-1+i & -1-i\\
	1-i & 1+i
	\end{pmatrix}
	=-I_{1},
	\end{align}
	and the following $B_{ij}$ matrices%
	\begin{align}
	B_{11}  &  =%
	\begin{pmatrix}
	1 & 0\\
	0 & 0
	\end{pmatrix}
	,B_{21}=%
	\begin{pmatrix}
	0 & 0\\
	0 & 1
	\end{pmatrix}
	,\nonumber\\
	B_{12}  &  =%
	\begin{pmatrix}
	0 & 0\\
	0 & 1
	\end{pmatrix}
	,B_{22}=%
	\begin{pmatrix}
	1 & 0\\
	0 & 0
	\end{pmatrix}
	.
	\end{align}

	The complex ADHM equations can be verified to be%
	\begin{align}
	&  \left[  B_{11},B_{12}\right]  +I_{1}J_{1}\nonumber\\
	&  =%
	\begin{pmatrix}
	1 & 0\\
	0 & 0
	\end{pmatrix}%
	\begin{pmatrix}
	0 & 0\\
	0 & 1
	\end{pmatrix}
	-%
	\begin{pmatrix}
	0 & 0\\
	0 & 1
	\end{pmatrix}%
	\begin{pmatrix}
	1 & 0\\
	0 & 0
	\end{pmatrix}
	+%
	\begin{pmatrix}
	1-i & 1+i\\
	-1+i & -1-i
	\end{pmatrix}%
	\begin{pmatrix}
	1+i & -i-1\\
	-1+i & 1-i
	\end{pmatrix}
	\nonumber\\
	&  =%
	\begin{pmatrix}
	0 & 0\\
	0 & 0
	\end{pmatrix}
	-%
	\begin{pmatrix}
	0 & 0\\
	0 & 0
	\end{pmatrix}
	+%
	\begin{pmatrix}
	0 & 0\\
	0 & 0
	\end{pmatrix}
	=%
	\begin{pmatrix}
	0 & 0\\
	0 & 0
	\end{pmatrix}
	,
	\end{align}%
	\begin{align}
	&  \left[  B_{21},B_{22}\right]  +I_{2}J_{2}=\left[  B_{21},B_{22}\right]
	+I_{1}J_{1}\nonumber\\
	&  =%
	\begin{pmatrix}
	0 & 0\\
	0 & 1
	\end{pmatrix}%
	\begin{pmatrix}
	1 & 0\\
	0 & 0
	\end{pmatrix}
	-%
	\begin{pmatrix}
	1 & 0\\
	0 & 0
	\end{pmatrix}%
	\begin{pmatrix}
	0 & 0\\
	0 & 1
	\end{pmatrix}
	+%
	\begin{pmatrix}
	0 & 0\\
	0 & 0
	\end{pmatrix}
	\nonumber\\
	&  =%
	\begin{pmatrix}
	0 & 0\\
	0 & 0
	\end{pmatrix}
	-%
	\begin{pmatrix}
	0 & 0\\
	0 & 0
	\end{pmatrix}
	+%
	\begin{pmatrix}
	0 & 0\\
	0 & 0
	\end{pmatrix}
	=%
	\begin{pmatrix}
	0 & 0\\
	0 & 0
	\end{pmatrix}
	,
	\end{align}
	and%
	\begin{align}
	&  [B_{11},B_{22}]+\left[  B_{21},B_{12}\right]  +I_{1}.J_{2}+I_{2}%
	J_{1}=[B_{11},B_{22}]+\left[  B_{21},B_{12}\right]  -I_{1}.J_{1}-I_{1}%
	J_{1}\nonumber\\
	&  =%
	\begin{pmatrix}
	1 & 0\\
	0 & 0
	\end{pmatrix}%
	\begin{pmatrix}
	1 & 0\\
	0 & 0
	\end{pmatrix}
	-%
	\begin{pmatrix}
	1 & 0\\
	0 & 0
	\end{pmatrix}%
	\begin{pmatrix}
	1 & 0\\
	0 & 0
	\end{pmatrix}
	+%
	\begin{pmatrix}
	0 & 0\\
	0 & 1
	\end{pmatrix}%
	\begin{pmatrix}
	0 & 0\\
	0 & 1
	\end{pmatrix}
	-%
	\begin{pmatrix}
	0 & 0\\
	0 & 1
	\end{pmatrix}%
	\begin{pmatrix}
	0 & 0\\
	0 & 1
	\end{pmatrix}
	\nonumber\\
	&  -%
	\begin{pmatrix}
	0 & 0\\
	0 & 0
	\end{pmatrix}
	-%
	\begin{pmatrix}
	0 & 0\\
	0 & 0
	\end{pmatrix}
	=%
	\begin{pmatrix}
	0 & 0\\
	0 & 0
	\end{pmatrix}
	.
	\end{align}

	\bigskip Finally we need to check the following equations%
	\begin{align}
	zB_{11}+wB_{21}+x &  =%
	\begin{pmatrix}
	z+x & 0\\
	0 & w+x
	\end{pmatrix}
	=0,\\
	zB_{12}+wB_{22}+y &  =%
	\begin{pmatrix}
	w+y & 0\\
	0 & z+y
	\end{pmatrix}
	=0.
	\end{align}
	Since we want $\frac{w}{z}=1$, we obtain%
	\begin{equation}
	w=z,x=-z,y=-z,
	\end{equation}
	which represents a point on $CP^{3}$%
	\begin{equation}
	\lbrack x:y:z:w]=[-1:-1:1:1].
	\end{equation}
	This is the sheaf point where jumping occurs.
	
	In the following, as an example, we will calculate explicitly the complex YM
	$2$-instanton field strength for the ADHM data given in Eq.(\ref{dat1}),
	Eq.(\ref{dat2}), Eq.(\ref{dat3}) and Eq.(\ref{dat4}). The $\Delta$ matrix in
	Eq.(\ref{ab}) is \cite{CSW}
	\begin{equation}
	\Delta=%
	\begin{pmatrix}
	\lambda_{1} & \lambda_{2}\\
	x+y_{11} & 0\\
	0 & x+y_{22}%
	\end{pmatrix}
	, \label{data}%
	\end{equation}
	and its biquaternion conjugation is \cite{Ann}%
	\begin{equation}
	\Delta^{\circledast}=%
	\begin{pmatrix}
	\lambda_{1}^{\circledast} & x^{\circledast}+y_{11}^{\circledast} & 0\\
	\lambda_{2}^{\circledast} & 0 & x^{\circledast}+y_{22}^{\circledast}%
	\end{pmatrix}
	.
	\end{equation}
	Note that $x^{\circledast}=x^{\dagger}$ as $x_{\mu}$ is real. The next step is
	to introduce%
	\begin{equation}
	v=%
	\begin{pmatrix}
	v_{1}\\
	v_{2}\\
	v_{3}%
	\end{pmatrix}
	\end{equation}
	which satisfies%
	\begin{equation}
	\Delta^{\circledast}v=0, \label{v1}%
	\end{equation}
	and the normalization condition%
	\begin{equation}
	v^{\circledast}v=1. \label{v2}%
	\end{equation}
	Eq.(\ref{v1}) can be written as%
	\begin{equation}%
	\begin{pmatrix}
	\lambda_{1}^{\circledast} & x^{\circledast}+y_{11}^{\circledast} & 0\\
	\lambda_{2}^{\circledast} & 0 & x^{\circledast}+y_{22}^{\circledast}%
	\end{pmatrix}%
	\begin{pmatrix}
	v_{1}\\
	v_{2}\\
	v_{3}%
	\end{pmatrix}
	=0
	\end{equation}
	or%
	\begin{align}
	\lambda_{1}^{\circledast}v_{1}+\left(  x^{\circledast}+y_{11}^{\circledast
	}\right)  v_{2}  &  =0,\\
	\lambda_{2}^{\circledast}v_{1}+\left(  x^{\circledast}+y_{22}^{\circledast
	}\right)  v_{3}  &  =0,
	\end{align}
	which can be solved to be%
	\begin{align}
	v_{2}  &  =\frac{-\left(  x+y_{11}\right)  \lambda_{1}^{\circledast}%
	}{\left\vert x+y_{11}\right\vert _{c}^{2}}v_{1},\\
	v_{3}  &  =\frac{-\left(  x+y_{22}\right)  \lambda_{2}^{\circledast}%
	}{\left\vert x+y_{22}\right\vert _{c}^{2}}v_{1}.
	\end{align}
	It is important to note that since $y_{12}=0$, the solvability of $v$ is
	greatly simplified.
	
	The normalization condition in Eq.(\ref{v2}) can now be written as
	\begin{equation}
	v_{1}^{\circledast}v_{1}+\frac{v_{1}^{\circledast}\lambda_{1}\left(
		x+y_{11}\right)  ^{\circledast}}{\left\vert x+y_{11}\right\vert _{c}^{2}}%
	\frac{\left(  x+y_{11}\right)  \lambda_{1}^{\circledast}v_{1}}{\left\vert
		x+y_{11}\right\vert _{c}^{2}}+\frac{v_{1}^{\circledast}\lambda_{2}\left(
		x+y_{22}\right)  ^{\circledast}}{\left\vert x+y_{22}\right\vert _{c}^{2}}%
	\frac{\left(  x+y_{22}\right)  \lambda_{2}^{\circledast}v_{1}}{\left\vert
		x+y_{22}\right\vert _{c}^{2}}=1
	\end{equation}
	or%
	\begin{equation}
	v_{1}^{\circledast}v_{1}\left[  1+\frac{\left\vert \lambda_{1}\right\vert
		_{c}^{2}}{\left\vert x+y_{11}\right\vert _{c}^{2}}+\frac{\left\vert
		\lambda_{2}\right\vert _{c}^{2}}{\left\vert x+y_{22}\right\vert _{c}^{2}%
	}\right]  =1.
	\end{equation}
	Note that
	\begin{align}
	\left\vert \lambda_{1}\right\vert _{c}^{2} &  =1^{2}+\left(  -i\right)
	^{2}+\left(  i\right)  ^{2}+\left(  -1\right)  ^{2}=0,\label{v3}\\
	\left\vert \lambda_{2}\right\vert _{c}^{2} &  =\left(  -1\right)  ^{2}+\left(
	i\right)  ^{2}+\left(  -i\right)  ^{2}+\left(  1\right)  ^{2}=0,\label{v4}%
	\end{align}
	so we have%
	\begin{equation}
	v_{1}^{\circledast}v_{1}=1
	\end{equation}
	For simplicity, we can choose the $v_{1}$ biquaternion to be a pure number%
	\begin{equation}
	v_{1}=1
	\end{equation}
	to get%
	\begin{equation}
	v=%
	\begin{pmatrix}
	v_{1}\\
	v_{2}\\
	v_{3}%
	\end{pmatrix}
	=%
	\begin{pmatrix}
	1\\
	\frac{-\left(  x+y_{11}\right)  \lambda_{1}^{\circledast}}{\left\vert
		x+y_{11}\right\vert _{c}^{2}}\\
	\frac{-\left(  x+y_{22}\right)  \lambda_{2}^{\circledast}}{\left\vert
		x+y_{22}\right\vert _{c}^{2}}%
	\end{pmatrix}
	.
	\end{equation}

	Let's now calculate $f^{-1}$ to be%
	\begin{align}
	f^{-1} &  =\Delta^{\circledast}\Delta=%
	\begin{pmatrix}
	\lambda_{1}^{\circledast} & x^{\circledast}+y_{11}^{\circledast} & 0\\
	\lambda_{2}^{\circledast} & 0 & x^{\circledast}+y_{22}^{\circledast}%
	\end{pmatrix}%
	\begin{pmatrix}
	\lambda_{1} & \lambda_{2}\\
	x+y_{11} & 0\\
	0 & x+y_{22}%
	\end{pmatrix}
	\nonumber\\
	&  =%
	\begin{pmatrix}
	\lambda_{1}^{\circledast}\lambda_{1}+\left\vert x+y_{11}\right\vert _{c}^{2} &
	\lambda_{1}^{\circledast}\lambda_{2}\\
	\lambda_{2}^{\circledast}\lambda_{1} & \lambda_{2}^{\circledast}\lambda
	_{2}+\left\vert x+y_{22}\right\vert _{c}^{2}%
	\end{pmatrix}
	\nonumber\\
	&  =%
	\begin{pmatrix}
	\left\vert x+y_{11}\right\vert _{c}^{2} & 0\\
	0 & \left\vert x+y_{22}\right\vert _{c}^{2}%
	\end{pmatrix}
	\label{ff}%
	\end{align}
	where we have used the results of Eq.(\ref{v3}) and Eq.(\ref{v4}) and the
	following amasing simple results%
	\begin{equation}
	\lambda_{1}^{\circledast}\lambda_{2}=\left(  e_{0}+ie_{1}-ie_{2}+e_{3}\right)
	\left(  -e_{0}+ie_{1}-ie_{2}+e_{3}\right)  =0,
	\end{equation}%
	\begin{equation}
	\lambda_{2}^{\circledast}\lambda_{1}=\left(  -e_{0}-ie_{1}+ie_{2}%
	-e_{3}\right)  \left(  e_{0}-ie_{1}+ie_{2}-e_{3}\right)  =0.
	\end{equation}
	So we have%
	\begin{equation}
	f=%
	\begin{pmatrix}
	\frac{1}{\left\vert x+y_{11}\right\vert _{c}^{2}} & 0\\
	0 & \frac{1}{\left\vert x+y_{22}\right\vert _{c}^{2}}%
	\end{pmatrix}
	.
	\end{equation}
	Finally one can explicitly calculate the complex YM $2$-instanton field
	strength to be%
	\begin{align}
	F_{\mu\nu} &  =v^{\circledast}b\left(  e_{\mu}e_{\nu}^{\dagger}-e_{\nu}e_{\mu
	}^{\dagger}\right)  fb^{\circledast}v\nonumber\\
	&  =%
	\begin{pmatrix}
	1 & -\frac{\lambda_{1}\left(  x+y_{11}\right)  ^{\circledast}}{\left\vert
		x+y_{11}\right\vert _{c}^{2}} & -\frac{\lambda_{2}\left(  x+y_{22}\right)
		^{\circledast}}{\left\vert x+y_{22}\right\vert _{c}^{2}}%
	\end{pmatrix}%
	\begin{pmatrix}
	0 & 0\\
	1 & 0\\
	0 & 1
	\end{pmatrix}
	\left(  e_{\mu}e_{\nu}^{\dagger}-e_{\nu}e_{\mu}^{\dagger}\right)  \nonumber\\
	&  \cdot%
	\begin{pmatrix}
	\frac{1}{\left\vert x+y_{11}\right\vert _{c}^{2}} & 0\\
	0 & \frac{1}{\left\vert x+y_{22}\right\vert _{c}^{2}}%
	\end{pmatrix}%
	\begin{pmatrix}
	0 & 1 & 0\\
	0 & 0 & 1
	\end{pmatrix}%
	\begin{pmatrix}
	1\\
	\frac{-\left(  x+y_{11}\right)  \lambda_{1}^{\circledast}}{\left\vert
		x+y_{11}\right\vert _{c}^{2}}\\
	\frac{-\left(  x+y_{22}\right)  \lambda_{2}^{\circledast}}{\left\vert
		x+y_{22}\right\vert _{c}^{2}}%
	\end{pmatrix}
	\nonumber\\
	&  =%
	\begin{pmatrix}
	-\frac{\lambda_{1}\left(  x+y_{11}\right)  ^{\circledast}}{\left\vert
		x+y_{11}\right\vert _{c}^{2}} & -\frac{\lambda_{2}\left(  x+y_{22}\right)
		^{\circledast}}{\left\vert x+y_{22}\right\vert _{c}^{2}}%
	\end{pmatrix}
	\left(  e_{\mu}e_{\nu}^{\dagger}-e_{\nu}e_{\mu}^{\dagger}\right)
	\begin{pmatrix}
	\frac{1}{\left\vert x+y_{11}\right\vert _{c}^{2}} & 0\\
	0 & \frac{1}{\left\vert x+y_{22}\right\vert _{c}^{2}}%
	\end{pmatrix}%
	\begin{pmatrix}
	\frac{-\left(  x+y_{11}\right)  \lambda_{1}^{\circledast}}{\left\vert
		x+y_{11}\right\vert _{c}^{2}}\\
	\frac{-\left(  x+y_{22}\right)  \lambda_{2}^{\circledast}}{\left\vert
		x+y_{22}\right\vert _{c}^{2}}%
	\end{pmatrix}
	\nonumber\\
	&  =\frac{\lambda_{1}\left(  x+y_{11}\right)  ^{\circledast}\left(  e_{\mu
		}e_{\nu}^{\dagger}-e_{\nu}e_{\mu}^{\dagger}\right)  \left(  x+y_{11}\right)
		\lambda_{1}^{\circledast}}{\left\vert x+y_{11}\right\vert _{c}^{6}}%
	+\frac{\lambda_{2}\left(  x+y_{22}\right)  ^{\circledast}\left(  e_{\mu}%
		e_{\nu}^{\dagger}-e_{\nu}e_{\mu}^{\dagger}\right)  \left(  x+y_{22}\right)
		\lambda_{2}^{\circledast}}{\left\vert x+y_{22}\right\vert _{c}^{6}}.\label{v5}%
	\end{align}

	The amazingly simple result of Eq.(\ref{v5}) is to be compared to the CFTW
	\cite{CFTW} real $2$-instanton solution \cite{1977} which is lengthy and quite
	complicated. Presumably, there are many simplified mechanisms \cite{Ann4} for
	the calculation of complex YM instanton sheaves which make the solvability
	available. Moreover, the two terms in Eq.(\ref{v5}) give the singular
	structure of the field strength at the sheaf point $[x:y:z:w]=[-1:-1:1:1]$ on
	$CP^{3}$ as expected \cite{Ann4}.
	
	Let's check this singular structure in the following. We will follow the
	calculation in \cite{Ann4}. First of all, we note that the Pl\"{u}cker
	coordinate of the real line in $CP^{3}$ corresponding to the sheaf point
	$[-1:-1:1:1]$ or in short \textit{sheaf line} \cite{Ann4} can be calculated to
	be%
	\begin{align}
	\lbrack-1  &  :-1:1:1]\wedge\sigma\lbrack-1:-1:1:1]=[-1:-1:1:1]\wedge
	\lbrack-1:1:1:-1]\nonumber\\
	&  =[-2:0:2:-2:0:-2]\simeq\lbrack1:0:-1:1:0:1]=[z_{12}:z_{13}:z_{14}%
	:z_{23}:z_{24}:z_{34}]
	\end{align}
	where the $\sigma$ map was defined to be \cite{Ann4}%
	\begin{equation}
	\sigma:[z_{1}:z_{2}:z_{3}:z_{4}]\rightarrow\lbrack\overline{z}_{2}%
	:-\overline{z}_{1}:\overline{z}_{4}:-\overline{z}_{3}].
	\end{equation}

	On the other hand, the projection of the sheaf point $[-1:-1:1:1]$ on $CP^{3}$
	down to $S^{4}$ is%
	\begin{align}
	x &  =x_{0}e_{0}+x_{1}e_{1}+x_{2}e_{2}+x_{3}e_{3}\nonumber\\
	&  =%
	\begin{pmatrix}
	x_{0}-ix_{3} & -\left(  x_{2}+ix_{1}\right)  \\
	x_{2}-ix_{1} & x_{0}+ix_{3}%
	\end{pmatrix}
	\nonumber\\
	&  =%
	\begin{pmatrix}
	x_{11} & x_{21}\\
	x_{12} & x_{22}%
	\end{pmatrix}
	\end{align}
	with the following identification \cite{Ann4}
	\begin{align}
	-x_{21} &  =z_{13},\nonumber\\
	x_{11} &  =z_{14},\nonumber\\
	-x_{22} &  =z_{23},\nonumber\\
	x_{12} &  =z_{24},\nonumber\\
	1 &  =z_{34}.
	\end{align}
	With the above inputs, one easily calculates the sheaf point on $CP^{3}$ down
	to $S^{4}$ to be%
	\begin{equation}
	(x_{0},x_{1},x_{2},x_{3})=(-1,0,0,0).\label{int}%
	\end{equation}
	Finally by using the data $y_{11}=e_{0}$ in Eq.(\ref{dat1}), we get%
	\begin{equation}
	x+y_{11}=-e_{0}+e_{0}=0,
	\end{equation}
	which gives the singular structure of the first term in Eq.(\ref{v5}). One
	notes that the sheaf point in Eq.(\ref{int}) is also a singular point of the
	second term in Eq.(\ref{v5})%
	\begin{equation}
	\left\vert x+y_{22}\right\vert _{c}^{2}=(-1)^{2}+(i)^{2}+0^{2}+0^{2}%
	=0.\label{22}%
	\end{equation}

	Indeed, the set of jumping lines $J$ defined in Eq.(\ref{points}) for the ADHM
	data in Eq.(\ref{data}) can be calculated to be%
	\begin{align}
	\det\Delta(x)^{\circledast}\Delta(x) &  =\det f^{-1}=\left\vert x+y_{11}%
	\right\vert _{c}^{2}\cdot\left\vert x+y_{11}\right\vert _{c}^{2}\nonumber\\
	&  =[(x_{0}+1)^{2}+x_{1}^{2}+x_{2}^{2}+x_{3}^{2}][x_{0}^{2}+(x_{1}%
	+i)^{2}+x_{2}^{2}+x_{3}^{2}]=0\label{zero}%
	\end{align}
	where we have used the result of Eq.(\ref{ff}). In section IV.C of our
	previous publication \cite{Ann4}, the order of singularity of a jumping line
	was defined to be the singularity in $f$ or structure of zero in
	Eq.(\ref{zero}). For the vanishing of the first factor in Eq.(\ref{zero})
	\begin{equation}
	(x_{0}+1)^{2}+x_{1}^{2}+x_{2}^{2}+x_{3}^{2}=0,
	\end{equation}
	we get the sheaf point in Eq.(\ref{int}). On the other hand, the real
	solutions of%
	\begin{equation}
	x_{0}^{2}+(x_{1}+i)^{2}+x_{2}^{2}+x_{3}^{2}=0
	\end{equation}
	are%
	\begin{equation}
	x_{1}=0\text{ \ },\text{ \ }x_{0}^{2}+x_{2}^{2}+x_{3}^{2}=1,\label{pp}%
	\end{equation}
	which is a unit $2$-sphere on the $x_{0}-x_{2}-x_{3}$ hyperplane. Note that
	Eq.(\ref{22}) implies the sheaf point in Eq.(\ref{int}) sits on the unit
	$2$-sphere in Eq.(\ref{pp}). This is consistent with our previous result that
	a sheaf line is always a jumping line \cite{Ann4}. Moreover, it is obvious
	that the order of singularity of $f$ at the sheaf point $x=(x_{0},x_{1}%
	,x_{2},x_{3})=(-1,0,0,0)$ is higher than those of other points on the
	$2$-sphere associated with normal jumping lines. This is again consistent with
	the conjecture made in \cite{Ann4}.
	
	\subsection{Sheaf case with rank$\beta=1$}
	
	\bigskip In this section, we consider another sheaf case with rank$\beta=1$ on
	some points of $CP^{3}$ for some given ADHM data. As in the rank$\beta=0$
	case, we are looking for the factorization of a $\det$ factor in the $y_{12}$
	matrix%
	\begin{align}
	y_{12}  &  =\frac{i}{\left\vert y_{11}-y_{22}\right\vert _{c}^{2}}%
	\begin{pmatrix}
	d_{1}+a_{1} & d_{2}+a_{2}\\
	d_{3}+a_{3} & d_{4}+a_{4}%
	\end{pmatrix}%
	\begin{pmatrix}
	l & m-in\\
	m+in & -l
	\end{pmatrix}
	\nonumber\\
	&  =\frac{i}{\det}%
	\begin{bmatrix}
	A & B\\
	\tilde{A} & \tilde{B}%
	\end{bmatrix}%
	\begin{bmatrix}
	a & b\\
	c & d
	\end{bmatrix}
	.
	\end{align}
	We will see that under some general assumption on the moduli parameters, the
	matrix on the rhs of the following equation%
	
	\begin{equation}%
	\begin{bmatrix}
	A & B\\
	\tilde{A} & \tilde{B}%
	\end{bmatrix}%
	\begin{bmatrix}
	a & b\\
	c & d
	\end{bmatrix}
	=%
	\begin{bmatrix}
	Aa+Bc & Ab+Bd\\
	\tilde{A}a+\tilde{B}c & \tilde{A}b+\tilde{B}d
	\end{bmatrix}
	\label{mm}%
	\end{equation}
	factors out a common $\det$ factor%
	\begin{equation}
	\det=\left\vert y_{11}-y_{22}\right\vert _{c}^{2}=A\tilde{B}-B\tilde{A}.
	\end{equation}

	Let's first define%
	\begin{equation}
	\frac{Aa+Bc}{Ab+Bd}=-w^{\prime} \label{sin}%
	\end{equation}
	where $w^{\prime}$ will be identified to be the $w$ coordinate in $CP^{3}$ in
	the later calculation. We begin the calculation by assuming%
	\begin{equation}
	Ab+Bd=\varepsilon\det\label{sin2}%
	\end{equation}
	where $\varepsilon$ is a finite number as $\det\rightarrow0.$ One can then
	solve $d$
	\begin{equation}
	d=\frac{\varepsilon\det-Ab}{B} \label{ddd}%
	\end{equation}
	to obtain%
	\begin{align}
	&  \tilde{A}b+\tilde{B}d\nonumber\\
	&  =\left(  Ab+Bd\right)  +b\left(  \tilde{A}-A\right)  +d\left(  \tilde
	{B}-B\right) \nonumber\\
	&  =\varepsilon\det+b\left(  \tilde{A}-A\right)  +\left(  \frac{\varepsilon
		\det-Ab}{B}\right)  \left(  \tilde{B}-B\right) \nonumber\\
	&  =\varepsilon\det\left(  1+\frac{\tilde{B}-B}{B}\right)  +\frac{b}{B}\left[
	\left(  \tilde{A}-A\right)  B-A\left(  \tilde{B}-B\right)  \right] \nonumber\\
	&  =\det\left[  \varepsilon\frac{\tilde{B}}{B}-\frac{b}{B}\right]  =\det
	\tilde{\varepsilon}%
	\end{align}
	where we have defined
	\begin{equation}
	\tilde{\varepsilon}=\varepsilon\frac{\tilde{B}}{B}-\frac{b}{B},
	\end{equation}
	which was assumed to be finite as $\det\rightarrow0.$
	
	We can factor out a $\det$ factor in $Aa+Bc$ by using Eq.(\ref{sin}) and
	Eq.(\ref{sin2}) to obtain
	\begin{equation}
	Aa+Bc=-w^{\prime}\varepsilon\det.
	\end{equation}

	Finally, let's calculate
	\begin{align}
	&  \tilde{A}a+\tilde{B}c\nonumber\\
	&  =\left(  \det\tilde{\varepsilon}\right)  \left(  \frac{\tilde{A}a+\tilde
		{B}c}{\tilde{A}b+\tilde{B}d}\right) \nonumber\\
	&  =\left(  \det\tilde{\varepsilon}\right)  \left(  \frac{Aa+Bc}{Ab+Bd}%
	-\frac{\left(  ad-bc\right)  \left(  A\tilde{B}-B\tilde{A}\right)  }{\left(
		\tilde{A}b+\tilde{B}d\right)  \left(  Ab+Bd\right)  }\right) \nonumber\\
	&  =\left(  -w^{\prime}\right)  \tilde{\varepsilon}\det-\left(  \frac
	{ad-bc}{\tilde{\varepsilon}\det}\frac{\det}{\varepsilon\det}\right)
	\tilde{\varepsilon}\det. \label{bb}%
	\end{align}
	The factor $ad-bc$ in Eq.(\ref{bb}) can be calculated to be
	\begin{align}
	&  ad-bc\nonumber\\
	&  =a\left(  \frac{\varepsilon\det-Ab}{B}\right)  -bc\nonumber\\
	&  =\frac{a\varepsilon}{B}\det-\frac{b}{B}\left(  Aa+Bc\right)  \label{aa}%
	\end{align}
	where we have used Eq.(\ref{ddd}). Finally%
	\begin{align}
	&  \tilde{A}a+\tilde{B}c=\left(  -w^{\prime}\right)  \tilde{\varepsilon}%
	\det-\left(  \frac{ad-bc}{\tilde{\varepsilon}\det}\frac{\det}{\varepsilon\det
	}\right)  \tilde{\varepsilon}\det\nonumber\\
	&  =-w^{\prime}\tilde{\varepsilon}\det-\frac{a}{B}\det+\frac{b}{\varepsilon
		B}\left(  Aa+Bc\right) \nonumber\\
	&  =\det\left(  -w^{\prime}\tilde{\varepsilon}-\frac{a}{B}\right)  +\frac
	{b}{\varepsilon B}\left(  -w^{\prime}\varepsilon\det\right) \nonumber\\
	&  =\left(  \delta-\frac{w^{\prime}b}{B}\right)  \det
	\end{align}
	where $\left(  \delta-\frac{w^{\prime}b}{B}\right)  $ is assumed to be finite
	as $\det\rightarrow0$ and $\delta$ is defined to be
	\begin{equation}
	\delta=-w^{\prime}\tilde{\varepsilon}-\frac{a}{B}. \label{del}%
	\end{equation}

	In sum, we have achieved the factorization%
	\begin{equation}%
	\begin{bmatrix}
	A & B\\
	\tilde{A} & \tilde{B}%
	\end{bmatrix}%
	\begin{bmatrix}
	a & b\\
	c & d
	\end{bmatrix}
	=%
	\begin{bmatrix}
	-w^{\prime}\varepsilon\det & \varepsilon\det\\
	\left(  \delta-\frac{w^{\prime}b}{B}\right)  \det & \tilde{\varepsilon}\det
	\end{bmatrix}
	.
	\end{equation}
	We can now calculate the finite $y_{12}$ in the $\det\rightarrow0$ limit to be%
	\begin{align}
	y_{12}  &  =\frac{i}{\det}%
	\begin{bmatrix}
	A & B\\
	\tilde{A} & \tilde{B}%
	\end{bmatrix}%
	\begin{bmatrix}
	a & b\\
	c & d
	\end{bmatrix}
	=\frac{i}{\det}%
	\begin{bmatrix}
	-w^{\prime}\varepsilon\det & \varepsilon\det\\
	\left(  \delta-\frac{w^{\prime}b}{B}\right)  \det & \tilde{\varepsilon}\det
	\end{bmatrix}
	\nonumber\\
	&  =i%
	\begin{bmatrix}
	-w^{\prime}\varepsilon & \varepsilon\\
	\left(  \delta-\frac{w^{\prime}b}{B}\right)  & \tilde{\varepsilon}%
	\end{bmatrix}
	=%
	\begin{bmatrix}
	c_{1} & c_{2}\\
	c_{3} & c_{4}%
	\end{bmatrix}
	. \label{ccc}%
	\end{align}
	On the other hand, if we assume%
	\begin{equation}
	\left(  \delta-\frac{w^{\prime}b}{B}\right)  =-w^{\prime}\tilde{\varepsilon}
	\label{del2}%
	\end{equation}
	we get%
	\begin{equation}%
	\begin{pmatrix}
	c_{1}\\
	c_{3}%
	\end{pmatrix}
	+w^{\prime}%
	\begin{pmatrix}
	c_{2}\\
	c_{4}%
	\end{pmatrix}
	=0. \label{c1c}%
	\end{equation}
	After identifying $\delta$ in Eq.(\ref{del}) and Eq.(\ref{del2}), we obtain%
	\begin{equation}
	w^{\prime}=\frac{-a}{b}. \label{abc}%
	\end{equation}

	We are now ready to check the rank$\beta=1$ condition. For simplicity, we take
	$z=1$ to obtain
	\begin{align}
	I_{1}+wI_{2} &  =%
	\begin{pmatrix}
	-\lambda_{1}^{2}+i\lambda_{1}^{1}-w\left(  \lambda_{1}^{0}+i\lambda_{1}%
	^{3}\right)   & \lambda_{1}^{0}-i\lambda_{1}^{3}-w\left(  \lambda_{1}%
	^{2}+i\lambda_{1}^{1}\right)  \\
	-\lambda_{2}^{2}+i\lambda_{2}^{1}-w\left(  \lambda_{2}^{0}+i\lambda_{2}%
	^{3}\right)   & \lambda_{2}^{0}-i\lambda_{2}^{3}-w\left(  \lambda_{2}%
	^{2}+i\lambda_{2}^{1}\right)
	\end{pmatrix}
	,\label{88}\\
	B_{11}+wB_{21}+x &  =%
	\begin{pmatrix}
	d_{1}+d_{2}w+x & c_{1}+c_{2}w\\
	c_{1}+c_{2}w & -a_{1}-a_{2}w+x
	\end{pmatrix}
	,\label{89}\\
	B_{12}+wB_{22}+y &  =%
	\begin{pmatrix}
	d_{3}+d_{4}w+y & c_{3}+c_{4}w\\
	c_{3}+c_{4}w & -a_{3}-a_{4}w+y
	\end{pmatrix}
	.\label{90}%
	\end{align}
	We assume a common eigenvector to be%
	\begin{equation}
	v=(1,0)
	\end{equation}
	in Eq.(\ref{d}), and take the following particular solutions\bigskip
	\ (vanishing of the first rows of matrices in Eq.(\ref{88}) to Eq.(\ref{90}))
	\begin{align}
	-\lambda_{1}^{2}+i\lambda_{1}^{1}-w\left(  \lambda_{1}^{0}+i\lambda_{1}%
	^{3}\right)   &  =0,\label{1}\\
	\lambda_{1}^{0}-i\lambda_{1}^{3}-w\left(  \lambda_{1}^{2}+i\lambda_{1}%
	^{1}\right)   &  =0\label{2}\\
	d_{1}+d_{2}w+x &  =0,\label{3}\\
	c_{1}+c_{2}w &  =0,\label{4}\\
	d_{3}+d_{4}w+y &  =0,\label{55}\\
	c_{3}+c_{4}w &  =0,\label{6}%
	\end{align}
	To solve Eq.(\ref{4}) and Eq.(\ref{6}), we choose $\ w=w^{\prime}$ or%
	\begin{equation}
	w=w^{\prime}=-\frac{Aa+Bc}{Ab+Bd}=\frac{-a}{b},
	\end{equation}
	and make use of Eq.(\ref{c1c}). On the other hand, Eq.(\ref{3}) and
	Eq.(\ref{55}) give%
	\begin{equation}
	x=-d_{1}-d_{2}w,\text{ \ }y=-d_{3}-d_{4}w.\label{xy}%
	\end{equation}
	Finally Eq.(\ref{1}) and Eq.(\ref{2}) put constraints on the $\lambda$
	parameters. In addition, one needs to take into account the condition of
	finite $c_{j}$ in Eq.(\ref{ccc}).
	
	In the following, let's give one explicit example of rank$\beta=1$ and
	$\det=0$ case. We begin with the following ADHM data%
	\begin{equation}
	y_{11}=%
	\begin{pmatrix}
	d_{1} & d_{2}\\
	d_{3} & d_{4}%
	\end{pmatrix}
	=%
	\begin{pmatrix}
	0 & 1\\
	1 & 0
	\end{pmatrix}
	,y_{22}=-%
	\begin{pmatrix}
	a_{1} & a_{2}\\
	a_{3} & a_{4}%
	\end{pmatrix}
	=%
	\begin{pmatrix}
	-p & 0\\
	0 & -p
	\end{pmatrix}
	, \label{adhm11}%
	\end{equation}
	which give%
	\begin{equation}
	\det=p^{2}-1.
	\end{equation}
	Now Eq.(\ref{mm}) can be written as
	\begin{equation}%
	\begin{bmatrix}
	p & 1\\
	1 & p
	\end{bmatrix}%
	\begin{bmatrix}
	a & b\\
	c & -a
	\end{bmatrix}
	=%
	\begin{bmatrix}
	pa+c & pb-a\\
	a-pc & b-pa
	\end{bmatrix}
	.
	\end{equation}
	Then Eq.(\ref{sin}) and Eq.(\ref{abc}) imply%
	\begin{equation}
	\frac{pa+c}{pb-a}=\frac{a}{b},
	\end{equation}
	which gives $c=1-p.$ On the other hand, Eq.(\ref{sin2}) gives (take
	$\varepsilon=1$)%
	\begin{equation}
	pb-a=p^{2}-1.
	\end{equation}
	For simplicity, we choose $b=p-1$, then $a=1-p.$ We can now calculate $y_{12}$
	to be%
	\begin{align}
	y_{12}  &  =\frac{i}{p^{2}-1}%
	\begin{bmatrix}
	p & 1\\
	1 & p
	\end{bmatrix}%
	\begin{bmatrix}
	1-p & p-1\\
	1-p & p-1
	\end{bmatrix}
	\nonumber\\
	&  =\frac{i}{p^{2}-1}%
	\begin{bmatrix}
	-1 & 1\\
	-1 & 1
	\end{bmatrix}
	(p^{2}-1)=i%
	\begin{bmatrix}
	-1 & 1\\
	-1 & 1
	\end{bmatrix}
	. \label{faf2}%
	\end{align}

	We thus have achieved the factorization and have cancelled out the $\det
	=p^{2}-1$ factor in Eq.(\ref{faf2}). Moreover the parameters we introduced
	during the calculation of factorization can be calculated to be%
	\begin{equation}
	\varepsilon=1,\tilde{\varepsilon}=1,\delta=-3,
	\end{equation}
	which are finite as $\det\rightarrow0$.
	
	We will take the $p\rightarrow-1$ limit after the factorization. In this
	limit, we get%
	\begin{equation}%
	\begin{bmatrix}
	a & b\\
	c & -a
	\end{bmatrix}
	=%
	\begin{bmatrix}
	2 & -2\\
	2 & -2
	\end{bmatrix}
	=%
	\begin{pmatrix}
	l & m-in\\
	m+in & -l
	\end{pmatrix}
	,
	\end{equation}
	which give%
	\begin{align}
	l  &  =%
	\begin{vmatrix}
	\lambda_{1}^{0} & \lambda_{1}^{3}\\
	\lambda_{2}^{0} & \lambda_{2}^{3}%
	\end{vmatrix}
	-%
	\begin{vmatrix}
	\lambda_{1}^{1} & \lambda_{1}^{2}\\
	\lambda_{2}^{1} & \lambda_{2}^{2}%
	\end{vmatrix}
	=2,\label{da1}\\
	m  &  =%
	\begin{vmatrix}
	\lambda_{1}^{0} & \lambda_{1}^{1}\\
	\lambda_{2}^{0} & \lambda_{2}^{1}%
	\end{vmatrix}
	-%
	\begin{vmatrix}
	\lambda_{1}^{2} & \lambda_{1}^{3}\\
	\lambda_{2}^{2} & \lambda_{2}^{3}%
	\end{vmatrix}
	=0,\label{da2}\\
	n  &  =%
	\begin{vmatrix}
	\lambda_{1}^{0} & \lambda_{1}^{2}\\
	\lambda_{2}^{0} & \lambda_{2}^{2}%
	\end{vmatrix}
	-%
	\begin{vmatrix}
	\lambda_{1}^{3} & \lambda_{1}^{1}\\
	\lambda_{2}^{3} & \lambda_{2}^{1}%
	\end{vmatrix}
	=-2i. \label{da3}%
	\end{align}
	On the other hand, with $w=1$, Eq.(\ref{1}) and Eq.(\ref{2}) become
	\begin{align}
	-\lambda_{1}^{2}+i\lambda_{1}^{1}-\left(  \lambda_{1}^{0}+i\lambda_{1}%
	^{3}\right)   &  =0,\label{da4}\\
	\lambda_{1}^{0}-i\lambda_{1}^{3}-\left(  \lambda_{1}^{2}+i\lambda_{1}%
	^{1}\right)   &  =0. \label{da6}%
	\end{align}

	For illustration, we will find a particular solution of Eq.(\ref{da1}) to
	Eq.(\ref{da6}). First of all, we note that Eq.(\ref{da4}) and Eq.(\ref{da6})
	give%
	\begin{align}
	\lambda_{1}^{0}-i\lambda_{1}^{1}  &  =0,\label{ma}\\
	i\lambda_{1}^{3}+\lambda_{1}^{2}  &  =0. \label{ma2}%
	\end{align}
	We choose the particular solution of Eq.(\ref{ma}) and Eq.(\ref{ma2}) to be
	\begin{equation}
	\lambda_{1}^{0}=1,\lambda_{1}^{1}=-i. \label{adhm12}%
	\end{equation}

	To further simplify the calculation, we choose%
	\begin{equation}
	\lambda_{1}^{2}=0=\lambda_{1}^{3}\text{ and }\lambda_{2}^{0}=0=\lambda_{2}%
	^{1}. \label{adhm13}%
	\end{equation}
	With these choices, Eq.(\ref{da2}) is trivially satisfied and Eq.(\ref{da1})
	and Eq.(\ref{da3}) give%
	\begin{align}
	\lambda_{2}^{2}-i\lambda_{2}^{3}  &  =-2i,\label{ma3}\\
	\lambda_{2}^{3}+i\lambda_{2}^{2}  &  =2. \label{ma4}%
	\end{align}
	We see that Eq.(\ref{ma3}) and Eq.(\ref{ma4}) are similar, and for simplicity
	we choose one particular solution to be%
	\begin{equation}
	\lambda_{2}^{2}=-i,=\lambda_{2}^{3}=1. \label{adhm14}%
	\end{equation}
	Finally%
	\begin{equation}
	x=-d_{1}-d_{2}w=-1,\text{ \ }y=-d_{3}-d_{4}w=-1.
	\end{equation}

	We can now check that rank$\beta=1.$ Indeed the $\beta$ matrix can be
	explicitly calculated to be
	\begin{align}
	\beta &  =%
	\begin{bmatrix}
	-zB_{12}-wB_{22}-y & zB_{11}+wB_{21}+x & zI_{1}+wI_{2}%
	\end{bmatrix}
	\nonumber\\
	&  =%
	\begin{pmatrix}
	0 & 0 & 0 & 0 & 0 & 0\\
	0 & -2 & 0 & -2 & 0 & 0
	\end{pmatrix}
	.
	\end{align}

	We conclude that for the ADHM data given in Eq.(\ref{adhm11}), Eq.(\ref{faf2}%
	), Eq.(\ref{adhm12}), Eq.(\ref{adhm13}) and Eq.(\ref{adhm14}), rank$\beta=1$
	at point $[x:y:z:w]=[-1:-1:1:1]$ on $CP^{3}.$ So this is a solution of complex
	YM instanton sheaf. Note that this extended complex YM instanton sheaf
	solution was not considered in \cite{Ann2} since $\det=0.$
	
	\subsection{Bundle case with rank$\beta=2$}
	
	In this section, we consider the bundle solutions with rank$\beta=2$ on the
	whole $CP^{3}$ for some given ADHM data. We are again looking for the
	factorization of a $\det$ factor in the $y_{12}$ matrix. We will demonstrate
	the existence of \ a class of extended complex YM $2$-instanton solutions with
	rank$\beta=2$ and $\det=\left\vert y_{11}-y_{22}\right\vert _{c}^{2}=0$. We
	begin with the following ADHM data
	
	\bigskip%
	\begin{equation}
	y_{11}\equiv%
	\begin{pmatrix}
	d_{1} & d_{2}\\
	d_{3} & d_{4}%
	\end{pmatrix}
	=%
	\begin{pmatrix}
	1 & -i\\
	1 & -i
	\end{pmatrix}
	,y_{22}=-%
	\begin{pmatrix}
	a_{1} & a_{2}\\
	a_{3} & a_{4}%
	\end{pmatrix}
	=-%
	\begin{pmatrix}
	a_{1} & a_{2}\\
	a_{3} & -ia_{3}%
	\end{pmatrix}
	,
	\end{equation}
	with $a_{4}=-ia_{3}.$ The $\det$ can then be calculated to be
	\begin{equation}
	\det=\left\vert y_{11}-y_{22}\right\vert _{c}^{2}=-\left(  a_{3}+1\right)
	\left(  ia_{1}+a_{2}\right)  . \label{tss}%
	\end{equation}

	We will take $\left(  ia_{1}+a_{2}\right)  \rightarrow0$ limit after factoring
	out the $\det$ factor in $y_{12}$. The $\lambda$ parameters are chosen to be
	\begin{align}
	\lambda_{1}^{0}  &  =1+\varepsilon,\lambda_{1}^{1}=0,\lambda_{1}^{2}%
	=i,\lambda_{1}^{3}=\delta,\nonumber\\
	\lambda_{2}^{0}  &  =0,\lambda_{2}^{1}=1,\lambda_{2}^{2}=0,\lambda_{2}^{3}=0
	\end{align}
	where%
	\begin{align}
	\varepsilon &  =\frac{-a_{1}+ia_{2}}{2\left(  1+a_{1}\right)  }\rightarrow0,\\
	\delta &  =\frac{-a_{1}+ia_{2}}{2i\left(  1+a_{1}\right)  }\rightarrow0
	\end{align}
	in the $\left(  ia_{1}+a_{2}\right)  \rightarrow0$ limit. The $l$, $m$, $n$
	parameters defined in Eq.(\ref{lmn}) can be calculated to be%
	\begin{equation}
	l=i,m=1+\varepsilon,n=-\delta.
	\end{equation}

	We are now ready to calculate $y_{12}$%
	\begin{align}
	y_{12}  &  =\frac{i}{\left\vert y_{11}-y_{22}\right\vert _{c}^{2}}%
	\begin{pmatrix}
	1+a_{1} & -i+a_{2}\\
	1+a_{3} & -i-ia_{3}%
	\end{pmatrix}%
	\begin{pmatrix}
	l & m-in\\
	m+in & -l
	\end{pmatrix}
	\nonumber\\
	&  =\frac{i}{\left\vert y_{11}-y_{22}\right\vert _{c}^{2}}%
	\begin{pmatrix}
	1+a_{1} & -i+a_{2}\\
	1+a_{3} & -i-ia_{3}%
	\end{pmatrix}%
	\begin{pmatrix}
	i & 1+\varepsilon+i\delta\\
	1+\varepsilon-i\delta & -i
	\end{pmatrix}
	\nonumber\\
	&  =\frac{i}{\left\vert y_{11}-y_{22}\right\vert _{c}^{2}}%
	\begin{pmatrix}
	1+a_{1} & -i+a_{2}\\
	1+a_{3} & -i+a_{4}%
	\end{pmatrix}%
	\begin{pmatrix}
	i & \frac{1+ia_{2}}{1+a_{1}}\\
	1 & -i
	\end{pmatrix}
	\end{align}
	where we have used%
	\begin{align}
	1+\varepsilon+i\delta &  =\frac{2\left(  1+a_{1}\right)  -a_{1}+ia_{2}+\left(
		-a_{1}+ia_{2}\right)  }{2\left(  1+a_{1}\right)  }=\frac{1+ia_{2}}{1+a_{1}},\\
	1+\varepsilon-i\delta &  =\frac{2\left(  1+a_{1}\right)  -a_{1}+ia_{2}-\left(
		-a_{1}+ia_{2}\right)  }{2\left(  1+a_{1}\right)  }=1.
	\end{align}
	The parameter $y_{12}$ can be further calculated to be%
	\begin{align}
	y_{12}  &  =\frac{i}{\left\vert y_{11}-y_{22}\right\vert _{c}^{2}}%
	\begin{pmatrix}
	ia_{1}+a_{2} & 0\\
	0 & \frac{\left(  1+a_{3}\right)  \left(  1+ia_{2}\right)  +\left(
		-i+a_{4}\right)  \left(  1+a_{1}\right)  }{1+a_{1}}%
	\end{pmatrix}
	\nonumber\\
	&  =\frac{i}{\left\vert y_{11}-y_{22}\right\vert _{c}^{2}}%
	\begin{pmatrix}
	\frac{-\left\vert y_{11}-y_{22}\right\vert _{c}^{2}}{1+a_{3}} & 0\\
	0 & \frac{-i\left\vert y_{11}-y_{22}\right\vert _{c}^{2}}{1+a_{1}}%
	\end{pmatrix}
	\label{out}%
	\end{align}
	where Eq.(\ref{tss}) has been used. One can now factor out the $\det$ factor
	in Eq.(\ref{out}) and take the $\det\rightarrow0$ limit to get%
	\begin{equation}
	y_{12}=%
	\begin{pmatrix}
	\frac{-i}{1+a_{3}} & 0\\
	0 & \frac{1}{1+a_{1}}%
	\end{pmatrix}
	.
	\end{equation}
	So the $y$ data is given by%
	\begin{align}%
	\begin{pmatrix}
	y_{11} & y_{12}\\
	y_{12} & y_{22}%
	\end{pmatrix}
	&  =%
	\begin{pmatrix}%
	\begin{pmatrix}
	1 & -i\\
	1 & -i
	\end{pmatrix}
	&
	\begin{pmatrix}
	\frac{-i}{1+a_{3}} & 0\\
	0 & \frac{1}{1+a_{1}}%
	\end{pmatrix}
	\\%
	\begin{pmatrix}
	\frac{-i}{1+a_{3}} & 0\\
	0 & \frac{1}{1+a_{1}}%
	\end{pmatrix}
	& -%
	\begin{pmatrix}
	a_{1} & a_{2}\\
	a_{3} & -ia_{3}%
	\end{pmatrix}
	\end{pmatrix}
	\nonumber\\
	&  =%
	\begin{pmatrix}
	1 & -i & \frac{-i}{1+a_{3}} & 0\\
	1 & -i & 0 & \frac{1}{1+a_{1}}\\
	\frac{-i}{1+a_{3}} & 0 & -a_{1} & -a_{2}\\
	0 & \frac{1}{1+a_{1}} & -a_{3} & ia_{3}%
	\end{pmatrix}
	.
	\end{align}
	After the rearrangement, we obtain the $B_{ij}$ matrix%
	\begin{align}
	B_{11}  &  =%
	\begin{pmatrix}
	1 & \frac{-i}{1+a_{3}}\\
	\frac{-i}{1+a_{3}} & -a_{1}%
	\end{pmatrix}
	,B_{21}=%
	\begin{pmatrix}
	-i & 0\\
	0 & -a_{2}%
	\end{pmatrix}
	,\\
	B_{12}  &  =%
	\begin{pmatrix}
	1 & 0\\
	0 & -a_{3}%
	\end{pmatrix}
	,B_{22}=%
	\begin{pmatrix}
	-i & \frac{1}{1+a_{1}}\\
	\frac{1}{1+a_{1}} & ia_{3}%
	\end{pmatrix}
	.
	\end{align}

	The next step is to check the rank$\beta=2$ condition. For this we will divide
	the whole $CP^{3}$ into three parts.
	
	(1) For the first case, we set $z=1$. The first two components of the $\beta$
	matrix can be calculated to be%
	\begin{align}
	&  B_{11}+wB_{21}+x\nonumber\\
	&  =%
	\begin{pmatrix}
	1-iw+x & \frac{-i}{1+a_{3}}\\
	\frac{-i}{1+a_{3}} & -a_{1}-a_{2}w+x
	\end{pmatrix}
	, \label{b1b}%
	\end{align}%
	\begin{align}
	&  B_{12}+wB_{22}+y\nonumber\\
	&  =%
	\begin{pmatrix}
	1-iw+y & \frac{1}{1+a_{1}}w\\
	\frac{1}{1+a_{1}}w & -a_{3}+ia_{3}w+y
	\end{pmatrix}
	, \label{b2b}%
	\end{align}
	and the last component
	\begin{equation}
	I_{1}+wI_{2}=%
	\begin{pmatrix}
	-\lambda_{1}^{2}+i\lambda_{1}^{1}-w\left(  \lambda_{1}^{0}+i\lambda_{1}%
	^{3}\right)  & \lambda_{1}^{0}-i\lambda_{1}^{3}-w\left(  \lambda_{1}%
	^{2}+i\lambda_{1}^{1}\right) \\
	-\lambda_{2}^{2}+i\lambda_{2}^{1}-w\left(  \lambda_{2}^{0}+i\lambda_{2}%
	^{3}\right)  & \lambda_{2}^{0}-i\lambda_{2}^{3}-w\left(  \lambda_{2}%
	^{2}+i\lambda_{2}^{1}\right)
	\end{pmatrix}
	=%
	\begin{pmatrix}
	-i-w & 1-wi\\
	i & -iw
	\end{pmatrix}
	\label{i1i}%
	\end{equation}

	Let's assume there exists sheaf solution for this case. We will see soon that
	a contradiction results. We first note that one necessary condition to have
	sheaf solution is%
	\begin{equation}
	\det\left(  I_{1}+wI_{2}\right)  =0,
	\end{equation}
	which can be written as%
	\begin{equation}
	\left(  n+im\right)  w^{2}+2ilw+\left(  n-im\right)  =0. \label{lmm}%
	\end{equation}
	The solution of Eq.(\ref{lmm}) in the $\det\rightarrow0$ limit is%
	
	\begin{align}
	w=\frac{1}{n+im}\left[  -il\pm\sqrt{-\left(  l^{2}+m^{2}+n^{2}\right)
	}\right]   & \nonumber\\
	=\frac{-i\pm\left(  -i\right)  \sqrt{-\left(  2\varepsilon+\varepsilon
			^{2}+\delta^{2}\right)  }}{1+\varepsilon+i\delta}\rightarrow-i  &  .
	\end{align}
	For this case, Eq.(\ref{b1b}) and Eq.(\ref{b2b}) reduce to
	\begin{align}
	B_{11}+wB_{21}+x  &  =%
	\begin{pmatrix}
	x & \frac{-i}{1+a_{3}}\\
	\frac{-i}{1+a_{3}} & x
	\end{pmatrix}
	,\\
	B_{12}+wB_{22}+y  &  =%
	\begin{pmatrix}
	y & \frac{-i}{1+a_{1}}\\
	\frac{-i}{1+a_{1}} & y
	\end{pmatrix}
	,
	\end{align}
	and Eq.(\ref{i1i}) reduces to
	\begin{equation}
	I_{1}+wI_{2}=%
	\begin{pmatrix}
	0 & 0\\
	i & -1
	\end{pmatrix}
	.
	\end{equation}
	In the end, the $\beta$ matrix can be written as%
	\begin{align}
	\beta &  =%
	\begin{bmatrix}
	-zB_{12}-wB_{22}-y & zB_{11}+wB_{21}+x & zI_{1}+wI_{2}%
	\end{bmatrix}
	\nonumber\\
	&  =%
	\begin{pmatrix}
	-y & \frac{i}{1+a_{1}} & x & \frac{-i}{1+a_{3}} & 0 & 0\\
	\frac{i}{1+a_{1}} & -y & \frac{-i}{1+a_{3}} & x & i & -1
	\end{pmatrix}
	,
	\end{align}
	which gives rank$\beta=2$ for any $x$ and $y$. We conclude that there exists
	no sheaf solution for this case. So this is an extended solution in bundle case.
	
	\bigskip(2) For the second case, we set $z=0$, $w=1$. The $\beta$ matrix can
	be calculated to be
	\begin{equation}
	zB_{11}+wB_{21}+x=%
	\begin{pmatrix}
	-i+x & 0\\
	0 & -a_{2}+x
	\end{pmatrix}
	,
	\end{equation}%
	\begin{equation}
	zB_{12}+wB_{22}+y=%
	\begin{pmatrix}
	-i+y & \frac{1}{1+a_{1}}\\
	\frac{1}{1+a_{1}} & ia_{3}+y
	\end{pmatrix}
	\end{equation}
	and%
	\begin{equation}
	zI_{1}+wI_{2}=%
	\begin{pmatrix}
	-1 & -i\\
	0 & -i
	\end{pmatrix}
	.
	\end{equation}
	So we get%
	\begin{align}
	\beta &  =%
	\begin{bmatrix}
	-zB_{12}-wB_{22}-y & zB_{11}+wB_{21}+x & zI_{1}+wI_{2}%
	\end{bmatrix}
	\nonumber\\
	&
	\begin{pmatrix}
	i-y & -\frac{1}{1+a_{1}} & -i+x & 0 & -1 & -i\\
	-\frac{1}{1+a_{1}} & -ia_{3}-y & 0 & -a_{2}+x & 0 & -i
	\end{pmatrix}
	,
	\end{align}
	which gives rank$\beta=2$, again a bundle case.
	
	(3) For the third case, we set $z=0$, $w=0$. The $\beta$ matrix can be
	calculated to be%
	\begin{align}
	&  zB_{11}+wB_{21}+x=%
	\begin{pmatrix}
	x & 0\\
	0 & x
	\end{pmatrix}
	,\\
	&  zB_{12}+wB_{22}+y=%
	\begin{pmatrix}
	y & 0\\
	0 & y
	\end{pmatrix}
	,\\
	&  zI_{1}+wI_{2}=%
	\begin{pmatrix}
	0 & 0\\
	0 & 0
	\end{pmatrix}
	.
	\end{align}
	So we get%
	\begin{align}
	\beta &  =%
	\begin{bmatrix}
	-zB_{12}-wB_{22}-y & zB_{11}+wB_{21}+x & zI_{1}+wI_{2}%
	\end{bmatrix}
	\nonumber\\
	&  =%
	\begin{pmatrix}
	-y & 0 & x & 0 & 0 & 0\\
	0 & -y & 0 & x & 0 & 0
	\end{pmatrix}
	,
	\end{align}
	which gives rank$\beta=2$ ($x$, $y$ can't be both zeros), again a bundle case.
	
	In this subsection, we thus have explicitly demonstrated a class of extended
	complex YM \ $2$-instanton solution with rank$\beta=2$ on the whole $CP^{3}$.
	
	\section{Conclusion}
	
	In this paper, we first show that the complex YM $2$-instanton sheaf with
	rank$\beta$= $0$ does not exist for the previous construction of $SU(2)$
	complex YM instantons with $\det=\left\vert y_{11}-y_{22}\right\vert _{c}%
	^{2}\neq0$ \cite{Ann}. The reason has been that complex YM $2$-instanton with
	rank$\beta=0$ implies $\det=\left\vert y_{11}-y_{22}\right\vert _{c}^{2}=0$
	which, although is allowed and possibly exist in the construction of complex
	YM $2$-instanton solutions, was not considered previously \cite{Ann}. We then
	proceed to show the existence of the new extended (or $\det=0$) complex YM
	$2$-instantons in this paper.
	
	Moreover, we discovered that the rank$\beta$ of these new extended complex YM
	instantons can be either $2$ on the whole $CP^{3}$ for some given ADHM data
	(bundle) or $1,$ $0$ on some points of $CP^{3}$ with some other given ADHM
	data (sheaves). We have also calculated explicit examples of these new
	instanton solutions with various rank$\beta$. These extended $SU(2)$ complex
	YM instantons have no real instanton counterparts.
	
	One unexpected result we obtained in the search of extended complex YM
	instantons was the discovery of the existence of instanton sheaf structure
	with diagonal $y$ ADHM data. For the $\det\neq0$ complex YM instantons
	constructed previously \cite{Ann}, it was shown \cite{Ann2} that there is no
	instanton sheaf solutions with diagonal $y$ ADHM data.
	
	It will be a challenge to generalize the calculation of $2$-instanton
	solutions presented in this paper to $k$-instanton solutions with higher
	topological charges. Another interesting issue is to check the solvability of
	the field strength $F$ of the new complex YM $2$-instanton sheaves discovered
	in this paper. The first explicit example of complex YM $2$-instanton field
	strength corresponding to sheaf case with rank$\beta=1$ and $\det\neq0$ was
	given in \cite{Ann4}. In this paper, we give the second example of complex YM
	$2$-instanton field strength in Eq.(\ref{v5}). This second example corresponds
	to the sheaf case with rank$\beta=0$ and $\det=0$. The "simple" explicit form
	of $2$-instanton field strength seems not available in the literature for the
	real instanton case \cite{1977}.
	
	\begin{acknowledgments}
		The work of J.C. Lee is supported in part by the Ministry of Science and
		Technology and S.T. Yau center of NCTU, Taiwan. The work of I-H. Tsai is
		supported by the Ministry of Science and Technology of Taiwan under grant
		number 106-2821-C-002-001-ES.
	\end{acknowledgments}

\end{document}